%
%
%
\def\unredoffs{} \def\redoffs{\voffset=-0.31truein\hoffset=0.125truein}
\def\speclscape{\special{}}
%
%
%
%
\newbox\leftpage \newdimen\fullhsize \newdimen\hstitle \newdimen\hsbody
\tolerance=1000\hfuzz=2pt
\catcode`\@=11 
\def\bigans{b }
\def\answ{b }
\ifx\answ\bigans\message{(This will come out unreduced.}
\magnification=\magstephalf\unredoffs\baselineskip=16pt plus 2pt minus 1pt
\hsbody=\hsize \hstitle=\hsize 
\else\message{(This will be reduced.} \let\l@r=L
\magnification=1000\baselineskip=16pt plus 2pt minus 1pt \vsize=7truein
\redoffs \hstitle=8truein\hsbody=4.75truein\fullhsize=10truein\hsize=\hsbody
\output={\ifnum\pageno=0 
  \shipout\vbox{\speclscape{\hsize\fullhsize\makeheadline}
    \hbox to \fullhsize{\hfill\pagebody\hfill}}\advancepageno
  \else
  \almostshipout{\leftline{\vbox{\pagebody\makefootline}}}\advancepageno 
  \fi}
\def\almostshipout#1{\if L\l@r \count1=1 \message{[\the\count0.\the\count1]}
      \global\setbox\leftpage=#1 \global\let\l@r=R
 \else \count1=2
  \shipout\vbox{\speclscape{\hsize\fullhsize\makeheadline}
      \hbox to\fullhsize{\box\leftpage\hfil#1}}  \global\let\l@r=L\fi}
\fi
%
\newcount\yearltd\yearltd=\year\advance\yearltd by -1900

\def\Title#1#2{\nopagenumbers\abstractfont\hsize=\hstitle\rightline{#1}%
\vskip 1in\centerline{\titlefont #2}\abstractfont\vskip .5in\pageno=0}
\def\Date#1{\vfill\leftline{#1}\tenpoint\supereject\global\hsize=\hsbody%
\footline={\hss\tenrm\folio\hss}}
%

\def\draftmode{\message{ DRAFTMODE }\def\draftdate{{\rm preliminary draft:
\number\month/\number\day/\number\yearltd\ \ \hourmin}}%
\headline={\hfil\draftdate}\writelabels\baselineskip=20pt plus 2pt minus 2pt
 {\count255=\time\divide\count255 by 60 \xdef\hourmin{\number\count255}
  \multiply\count255 by-60\advance\count255 by\time
  \xdef\hourmin{\hourmin:\ifnum\count255<10 0\fi\the\count255}}}
\def\nolabels{\def\wrlabeL##1{}\def\eqlabeL##1{}\def\reflabeL##1{}}
\def\writelabels{\def\wrlabeL##1{\leavevmode\vadjust{\rlap{\smash%
{\line{{\escapechar=` \hfill\rlap{\sevenrm\hskip.03in\string##1}}}}}}}%
\def\eqlabeL##1{{\escapechar-1\rlap{\sevenrm\hskip.05in\string##1}}}%
\def\reflabeL##1{\noexpand\llap{\noexpand\sevenrm\string\string\string##1}}}
\nolabels
%
\global\newcount\secno \global\secno=0
\global\newcount\meqno \global\meqno=1
\def\newsec#1{\global\advance\secno by1\message{(\the\secno. #1)}
\global\subsecno=0\eqnres@t\noindent{\bf\the\secno. #1}
\writetoca{{\secsym} {#1}}\par\nobreak\medskip\nobreak}
\def\eqnres@t{\xdef\secsym{\the\secno.}\global\meqno=1\bigbreak\bigskip}
\def\sequentialequations{\def\eqnres@t{\bigbreak}}\xdef\secsym{}
\global\newcount\subsecno \global\subsecno=0
\def\subsec#1{\global\advance\subsecno by1\message{(\secsym\the\subsecno. #1)}
\ifnum\lastpenalty>9000\else\bigbreak\fi
\noindent{\it\secsym\the\subsecno. #1}\writetoca{\string\quad 
{\secsym\the\subsecno.} {#1}}\par\nobreak\medskip\nobreak}
\def\appendix#1#2{\global\meqno=1\global\subsecno=0\xdef\secsym{\hbox{#1.}}
\bigbreak\bigskip\noindent{\bf Appendix #1. #2}\message{(#1. #2)}
\writetoca{Appendix {#1.} {#2}}\par\nobreak\medskip\nobreak}
%
%
\def\eqnn#1{\xdef #1{(\secsym\the\meqno)}\writedef{#1\leftbracket#1}%
\global\advance\meqno by1\wrlabeL#1}
\def\eqna#1{\xdef #1##1{\hbox{$(\secsym\the\meqno##1)$}}
\writedef{#1\numbersign1\leftbracket#1{\numbersign1}}%
\global\advance\meqno by1\wrlabeL{#1$\{\}$}}
\def\eqn#1#2{\xdef #1{(\secsym\the\meqno)}\writedef{#1\leftbracket#1}%
\global\advance\meqno by1$$#2\eqno#1\eqlabeL#1$$}
%
\newskip\footskip\footskip14pt plus 1pt minus 1pt 
\def\footnotefont{\ninepoint}\def\f@t#1{\footnotefont #1\@foot}
\def\f@@t{\baselineskip\footskip\bgroup\footnotefont\aftergroup\@foot\let\next}
\setbox\strutbox=\hbox{\vrule height9.5pt depth4.5pt width0pt}
\global\newcount\ftno \global\ftno=0
\def\foot{\global\advance\ftno by1\footnote{$^{\the\ftno}$}}
%
\newwrite\ftfile   
\def\footend{\def\foot{\global\advance\ftno by1\chardef\wfile=\ftfile
$^{\the\ftno}$\ifnum\ftno=1\immediate\openout\ftfile=foots.tmp\fi%
\immediate\write\ftfile{\noexpand\smallskip%
\noexpand\item{f\the\ftno:\ }\pctsign}\findarg}%
\def\footatend{\vfill\eject\immediate\closeout\ftfile{\parindent=20pt
\centerline{\bf Footnotes}\nobreak\bigskip\input foots.tmp }}}
\def\footatend{}
%
%
\global\newcount\refno \global\refno=1
\newwrite\rfile
\def\ref{[\the\refno]\nref}
\def\nref#1{\xdef#1{[\the\refno]}\writedef{#1\leftbracket#1}%
\ifnum\refno=1\immediate\openout\rfile=refs.tmp\fi
\global\advance\refno by1\chardef\wfile=\rfile\immediate
\write\rfile{\noexpand\item{#1\ }\reflabeL{#1\hskip.31in}\pctsign}\findarg}
\def\findarg#1#{\begingroup\obeylines\newlinechar=`\^^M\pass@rg}
{\obeylines\gdef\pass@rg#1{\writ@line\relax #1^^M\hbox{}^^M}%
\gdef\writ@line#1^^M{\expandafter\toks0\expandafter{\striprel@x #1}%
\edef\next{\the\toks0}\ifx\next\em@rk\let\next=\endgroup\else\ifx\next\empty%
\else\immediate\write\wfile{\the\toks0}\fi\let\next=\writ@line\fi\next\relax}}
\def\striprel@x#1{} \def\em@rk{\hbox{}} 
\def\lref{\begingroup\obeylines\lr@f}
\def\lr@f#1#2{\gdef#1{\ref#1{#2}}\endgroup\unskip}

\def\addref#1{\immediate\write\rfile{\noexpand\item{}#1}} 
\def\startrefs#1{\immediate\openout\rfile=refs.tmp\refno=#1}
\def\xref{\expandafter\xr@f}\def\xr@f[#1]{#1}
\def\refs#1{\count255=1[\r@fs #1{\hbox{}}]}
\def\r@fs#1{\ifx\und@fined#1\message{reflabel \string#1 is undefined.}%
\nref#1{need to supply reference \string#1.}\fi%
\vphantom{\hphantom{#1}}\edef\next{#1}\ifx\next\em@rk\def\next{}%
\else\ifx\next#1\ifodd\count255\relax\xref#1\count255=0\fi%
\else#1\count255=1\fi\let\next=\r@fs\fi\next}
%

%
\newwrite\ffile\global\newcount\figno \global\figno=1
\def\fig{fig.~\the\figno\nfig}
\def\nfig#1{\xdef#1{fig.~\the\figno}%
\writedef{#1\leftbracket fig.\noexpand~\the\figno}%
\ifnum\figno=1\immediate\openout\ffile=figs.tmp\fi\chardef\wfile=\ffile%
\immediate\write\ffile{\noexpand\medskip\noexpand\item{Fig.\ \the\figno. }
\reflabeL{#1\hskip.55in}\pctsign}\global\advance\figno by1\findarg}
\def\xfig{\expandafter\xf@g}\def\xf@g fig.\penalty\@M\ {}
\def\figs#1{figs.~\f@gs #1{\hbox{}}}
\def\f@gs#1{\edef\next{#1}\ifx\next\em@rk\def\next{}\else
\ifx\next#1\xfig #1\else#1\fi\let\next=\f@gs\fi\next}
\newwrite\lfile
{\escapechar-1\xdef\pctsign{\string\%}\xdef\leftbracket{\string\{}
\xdef\rightbracket{\string\}}\xdef\numbersign{\string\#}}

\def\writestop{\def\writestoppt{\immediate\write\lfile{\string\pageno%
\the\pageno\string\startrefs\leftbracket\the\refno\rightbracket%
\string\def\string\secsym\leftbracket\secsym\rightbracket%
\string\secno\the\secno\string\meqno\the\meqno}\immediate\closeout\lfile}}
\def\writestoppt{}\def\writedef#1{}
\def\seclab#1{\xdef #1{\the\secno}\writedef{#1\leftbracket#1}\wrlabeL{#1=#1}}
\def\subseclab#1{\xdef #1{\secsym\the\subsecno}%
\writedef{#1\leftbracket#1}\wrlabeL{#1=#1}}
\newwrite\tfile \def\writetoca#1{}
\def\leaderfill{\leaders\hbox to 1em{\hss.\hss}\hfill}
\def\writetoc{\immediate\openout\tfile=toc.tmp 
   \def\writetoca##1{{\edef\next{\write\tfile{\noindent ##1 
   \string\leaderfill {\noexpand\number\pageno} \par}}\next}}}

%
\catcode`\@=12 
%
\edef\tfontsize{\ifx\answ\bigans scaled\magstep3\else scaled\magstep4\fi}
\font\titlerm=cmr10 \tfontsize \font\titlerms=cmr7 \tfontsize
\font\titlermss=cmr5 \tfontsize \font\titlei=cmmi10 \tfontsize
\font\titleis=cmmi7 \tfontsize \font\titleiss=cmmi5 \tfontsize
\font\titlesy=cmsy10 \tfontsize \font\titlesys=cmsy7 \tfontsize
\font\titlesyss=cmsy5 \tfontsize \font\titleit=cmti10 \tfontsize
\skewchar\titlei='177 \skewchar\titleis='177 \skewchar\titleiss='177
\skewchar\titlesy='60 \skewchar\titlesys='60 \skewchar\titlesyss='60
\def\titlefont{\def\rm{\fam0\titlerm}
\textfont0=\titlerm \scriptfont0=\titlerms \scriptscriptfont0=\titlermss
\textfont1=\titlei \scriptfont1=\titleis \scriptscriptfont1=\titleiss
\textfont2=\titlesy \scriptfont2=\titlesys \scriptscriptfont2=\titlesyss
\textfont\itfam=\titleit \def\it{\fam\itfam\titleit}\rm}
 \ifx\answ\bigans\else scaled\magstep1\fi
\ifx\answ\bigans\def\abstractfont{\tenpoint}\else
\font\abssl=cmsl10 scaled \magstep1
\font\absrm=cmr10 scaled\magstep1 \font\absrms=cmr7 scaled\magstep1
\font\absrmss=cmr5 scaled\magstep1 \font\absi=cmmi10 scaled\magstep1
\font\absis=cmmi7 scaled\magstep1 \font\absiss=cmmi5 scaled\magstep1
\font\abssy=cmsy10 scaled\magstep1 \font\abssys=cmsy7 scaled\magstep1
\font\abssyss=cmsy5 scaled\magstep1 \font\absbf=cmbx10 scaled\magstep1
\skewchar\absi='177 \skewchar\absis='177 \skewchar\absiss='177
\skewchar\abssy='60 \skewchar\abssys='60 \skewchar\abssyss='60
\def\abstractfont{\def\rm{\fam0\absrm}
\textfont0=\absrm \scriptfont0=\absrms \scriptscriptfont0=\absrmss
\textfont1=\absi \scriptfont1=\absis \scriptscriptfont1=\absiss
\textfont2=\abssy \scriptfont2=\abssys \scriptscriptfont2=\abssyss
\textfont\itfam=\bigit \def\it{\fam\itfam\bigit}\def\footnotefont{\tenpoint}%
\textfont\slfam=\abssl \def\sl{\fam\slfam\abssl}%
\textfont\bffam=\absbf \def\bf{\fam\bffam\absbf}\rm}\fi
\def\tenpoint{\def\rm{\fam0\tenrm}
\textfont0=\tenrm \scriptfont0=\sevenrm \scriptscriptfont0=\fiverm
\textfont1=\teni  \scriptfont1=\seveni  \scriptscriptfont1=\fivei
\textfont2=\tensy \scriptfont2=\sevensy \scriptscriptfont2=\fivesy
\textfont\itfam=\tenit \def\it{\fam\itfam\tenit}\def\footnotefont{\ninepoint}%
\textfont\bffam=\tenbf \def\bf{\fam\bffam\tenbf}\def\sl{\fam\slfam\tensl}\rm}
\font\ninerm=cmr9 \font\sixrm=cmr6 \font\ninei=cmmi9 \font\sixi=cmmi6 
\font\ninesy=cmsy9 \font\sixsy=cmsy6 \font\ninebf=cmbx9 
\font\nineit=cmti9 \font\ninesl=cmsl9 \skewchar\ninei='177
\skewchar\sixi='177 \skewchar\ninesy='60 \skewchar\sixsy='60 
\def\ninepoint{\def\rm{\fam0\ninerm}
\textfont0=\ninerm \scriptfont0=\sixrm \scriptscriptfont0=\fiverm
\textfont1=\ninei \scriptfont1=\sixi \scriptscriptfont1=\fivei
\textfont2=\ninesy \scriptfont2=\sixsy \scriptscriptfont2=\fivesy
\textfont\itfam=\ninei \def\it{\fam\itfam\nineit}\def\sl{\fam\slfam\ninesl}%
\textfont\bffam=\ninebf \def\bf{\fam\bffam\ninebf}\rm} 
%
%

\hyphenation{anom-aly anom-alies coun-ter-term coun-ter-terms}
\def\inv{^{\raise.15ex\hbox{${\scriptscriptstyle -}$}\kern-.05em 1}}

\def\Dsl{\,\raise.15ex\hbox{/}\mkern-13.5mu D} 
\def\dsl{\raise.15ex\hbox{/}\kern-.57em\partial}

\font\bigit=cmti10 scaled \magstep1
\def\lspace{\ifx\answ\bigans{}\else\qquad\fi}
\def\lbspace{\ifx\answ\bigans{}\else\hskip-.2in\fi} 
\def\boxeqn#1{\vcenter{\vbox{\hrule\hbox{\vrule\kern3pt\vbox{\kern3pt
	\hbox{${\displaystyle #1}$}\kern3pt}\kern3pt\vrule}\hrule}}}
\def\mbox#1#2{\vcenter{\hrule \hbox{\vrule height#2in
		\kern#1in \vrule} \hrule}}  
%

\def\darr#1{\raise1.5ex\hbox{$\leftrightarrow$}\mkern-16.5mu #1}

\def\half{{\textstyle{1\over2}}} 
\def\roughly#1{\raise.3ex\hbox{$#1$\kern-.75em\lower1ex\hbox{$\sim$}}}
\input pictex
\def\half{{\textstyle{1\over 2}}}
\def\thalf{{\textstyle{3\over 2}}}
\def\fhalf{{\textstyle{5\over 2}}}
\def\neqno#1{\eqnn{#1} \eqno #1}
\def\stableau{\beginpicture
\setcoordinatesystem units <0.125truein,0.125truein>
\setplotarea x from 0 to 3, y from 0 to 1
\linethickness=1.0pt
\putrule from 0 0 to 3 0
\putrule from 0 1 to 3 1
\putrule from 0 0 to 0 1
\putrule from 1 0 to 1 1
\putrule from 2 0 to 2 1
\putrule from 3 0 to 3 1
\endpicture}
\def\atableau{\beginpicture
\setcoordinatesystem units <0.125truein,0.125truein>
\setplotarea x from 0 to 1, y from -1 to 2
\linethickness=1.0pt
\putrule from 0 -1 to 0 2
\putrule from 1 -1 to 1 2
\putrule from 0 -1 to 1 -1
\putrule from 0 0 to 1 0
\putrule from 0 1 to 1 1
\putrule from 0 2 to 1 2
\endpicture}
\def\mtableau{\beginpicture
\setcoordinatesystem units <0.125truein,0.125truein>
\setplotarea x from 0 to 2, y from -1 to 1
\linethickness=1.0pt
\putrule from 0 -1 to 0 1
\putrule from 1 -1 to 1 1
\putrule from 0 0 to 2 0
\putrule from 0 1 to 2 1
\putrule from 0 -1 to 1 -1
\putrule from 2 0 to 2 1
\endpicture}
%
%
%
\overfullrule=0pt

%

\Title{hep-ph/yymmdd, \#HUTP-98/A036}
{\vbox{\centerline{$S_3$ and the $L=1$ Baryons in }
\bigskip
\centerline{the Quark Model and the Chiral Quark Model}}}
\smallskip
\centerline{Hael Collins$^\dagger$   
       and Howard Georgi$^\ddagger$ } 
\smallskip
\centerline{\it Harvard University}
\centerline{\it Cambridge, MA 02138, USA}
\smallskip
\centerline{$^\dagger${\tt hael@feynman.harvard.edu} }
\centerline{$^\ddagger${\tt georgi@physics.harvard.edu} }

\bigskip

\medskip

\centerline{ABSTRACT}
\smallskip

\noindent
The $S_3$ symmetry corresponding to permuting the positions of the quarks
within a baryon allows us to study the $70$-plet of $L=1$ baryons without an
explicit choice for the spatial part of the quark wave functions: given a
set of operators with definite transformation properties under the
spin-flavor group $SU(3)\times SU(2)$ and under this $S_3$, the masses of
the baryons can be expressed in terms of a small number of unknown
parameters which are fit to the observed $L=1$ baryon mass spectrum.  This
approach is applied to study both the quark model and chiral constituent
quark model.  The latter theory leads to a set of mass perturbations which
more satisfactorily fits the observed $L=1$ baryon mass spectrum (though we
can say nothing, within our approach, about the physical reasonableness of
the parameters in the fit).  Predictions for the mixing angles and the
unobserved baryon masses are given for both models as well as a discussion
of specific baryons.

\Date{October, 1998}

\newsec{Introduction}

The non-relativistic quark model has been used extensively to study the
$L=1$ baryons \ref\isgurkarl{N.~Isgur and G.~Karl, ``$P$-wave Baryons in
the Quark Model,'' Phys.\ Rev.\ {\bf D18} (1978)
4187.}--\ref\dalitz{R.~R.~Horgan and R.~H.~Dalitz, ``Baryon Spectroscopy
and the Quark Shell Model (I) Nucl.\ Phys.\ {\bf B66} (1973) 135 and
Nucl.~Phys.~{\bf B71} (1974) 514, 546.  M.~Jones, R.~H.~Dalitz and
R.~R.~Horgan, ``Re-Analysis of the Baryon Mass Spectrum Using the Quark
Shell Model,'' Nucl.\ Phys.\ {\bf B129} (1977) 45.}.  In this model, the
observed mass spectrum of the baryons is generated by a two body Coulombic
potential \ref\dgg{A.~De R\' ujula, H.~Georgi and S.~L.~Glashow, ``Hadron
Masses in a Gauge Theory,'' Phys.\ Rev.\ {\bf D12} (1975) 147.}, produced
by a gluon exchange between two quarks.  The quark model leads to a
definite form for the $SU(3)\times SU(2)$ spin-flavor breaking interactions
but not for the ground state quark wave functions.  What is typically done
is to use harmonic oscillator wave functions for the spatial wave
functions.  The picture that then emerges for the baryon masses is
surprisingly good given the simplicity of the model; nevertheless, it has
several serious shortcomings, most notoriously its inability to explain the
lightness of the $\Lambda(1405)$.  Extensions of the quark model, such as
the inclusion of relativistic effects \ref\ic{S.~Capstick and N.~Isgur,
``Baryons in a relativized quark model with chromodynamics,'' Phys.\ Rev.\
{\bf D34} (1986) 2809.}\ although leading to a better agreement with the
entire spectrum of light mesons and baryons, do not seem to improve much
upon quark model's description of the $L=1$ baryons.

Another model has recently emerged to explain the observed baryon spectrum
\ref\glozman{L.~Ya.~Glozman, ``Baryons, Their Interactions and the Chiral 
Symmetry of QCD,'' Nucl.\ Phys.\ {\bf A629} (1998) 121; {\tt
hep-ph/9706361}.  F.~Stancu, S.~Pepin, L.~Ya.~Glozman, ``The
Nucleon-Nucleon Interaction in a Chiral Constituent Quark Model,'' Phys.\
Rev.\ {\bf C56} (1997) 2779; {\tt nucl-th/9705030}.}.  Its assumption is that
the correct effective theory within a baryon is not that of
constituent quarks exchanging gluons but rather that of the quasiparticles,
constituent quarks and Goldstone bosons, appropriate for energies below the
scale of chiral symmetry breaking.  The low energy quark potential in this
theory, which we refer to as the {\it chiral quark model\/}, is also a two
body Coulombic potential with an important difference---the inclusion of
flavor matrics at the quark-Goldstone boson vertices.  We shall here check
the claim that the different flavor structure of the chiral quark model
leads to a better fit with the observed $L=1$ spectrum.

The new ingredient in our study of the $L=1$ bosons is the use of the
permutation group $S_3$ to organize the spatial behavior of the quarks and
their interactions.  Physically, this symmetry corresponds to the fact that
the confining potential should treat the light quarks equivalently and
should be invariant under permutations of the positions of the three
quarks.  It allows us to circumvent choosing a definite form for the
quarks' spatial wave functions using group theory to keep track of our
ignorance of this spatial behavior.  What distinguishes one model from
another is the $SU(3)\times SU(2)$ spin-flavor and $S_3$ transformation
properties of the operators that produce the mass splittings among the
$L=1$ baryons.  We begin therefore with a description of this symmetry, how
the baryons transform under it and how it fits with the standard
$SU(3)\times SU(2)$ spin-flavor structure of the baryons and then proceed
to determine how the standard mass splitting operators transform under
$S_3$ in each of the models.

Our approach has a potential disadvantage, in addition to its obvious
advantages. Because we have not made dynamical assumptions about the wave
functions, but only used symmetry, we cannot say within our approach
whether the parameters of the fits we obtain are physically reasonable.
Thus our results should not be interpreted, by themselves, as evidence in
favor of the chiral quark model picture over the non-relativistic quark
model. However, we believe it is worth noting that difficulties for the
non-relativistic quark model persist even in this very general approach.

\newsec{$S_3$ and the $L=1$ Baryons.}

The $L=1$ negative parity baryons form a seventy dimensional representation
of the spin-flavor group $SU(6)$.  This 70-plet breaks into the
representations $^48$, $^28$, $^210$, $^21$ under separate spin and flavor
transformations; here the notation $^{2S+1}F$ indicates a multiplet that
forms an $F$ dimensional representation of the $SU(3)$ flavor group with
spin $S$.  Among the interactions that we shall consider are spin-orbit
couplings between quarks so that the baryonic states will be written,
and are measured, in terms of the total angular momentum, $J=L+S$.  The
baryonic states are then represented by linear superpostions of spatial,
flavor, spin, and orbital angular momentum wave functions for the three
contituent quarks.  The construction of these states for the $SU(3)\times
SU(2)$ spin-flavor part is straightforward in the non-relativistic limit,
but the spatial wave functions requires a specific dynamical model.  Some
of the earlier studies of the $L=1$ baryons used harmonic oscillator wave
functions for the spatial wave functions in terms of the relative positions
of the quarks (\isgurkarl\ and \dalitz).  One of the disadvantages of this
approach is that if a poor agreement is found for a model with the observed
baryon spectrum, it is not immediately clear whether that failure lies in
the model itself or in the specific choice of the spatial wave functions.

Fortunately, the spatial interactions of the quarks possess a symmetry
that allows us to escape the choice of a specific dynamical model for the
spatial wave functions.  The quark interactions should be invariant under
any permutation of the positions of the quarks.  This symmetry then implies
that the spatial wave functions should form a representation of the group
$S_3$.  We shall find that the $S_3$ group theory is sufficiently powerful
to reduce our ignorance of the spatial behavior to a small set of
constants.  The baryonic matrix elements will be linear combinations of
these $S_3$ constants whose coefficients depend upon the spin-flavor
assignments of the quark interactions and are completely determined by the
$SU(3)\times SU(2)$ group theory.  One of the advantages of this approach
is that it treats different models on the same footing.

The baryonic wave functions that we used are those used in
\ref\largen{C.~D.~Carone, H.~Georgi, L.~Kaplan, and D.~Morin, ``Decays of
$\ell=1$ Baryons---Quark Model versus Large-$N_c$,'' Phys.\ Rev.\ {\bf D50}
(1994) 5793--5807; {\tt hep-ph/9406227}.}\ in which each of the three-quark
states has only one of the quarks in an orbitally excited ($L=1$) state,
$|1,m\rangle$.  For example, the $|\Delta^{++};\half;\thalf,\thalf\rangle$
($|\Delta^{++};S;J,M_J\rangle$) state would thus be
$$\eqnn\deltaexp
\eqalignno{
|\Delta^{++};\half;\thalf,\thalf\rangle &=
- {\sqrt{2}\over 6}\,  uuu \Bigl\{
\psi^1_{11}(\vec r_1, \vec r_2, \vec r_3) 
\left( |\uparrow\downarrow\uparrow\rangle 
+ |\uparrow\uparrow\downarrow\rangle 
- 2 |\downarrow\uparrow\uparrow\rangle \right) &\cr
&\qquad\qquad\quad +
\psi^2_{11}(\vec r_1, \vec r_2, \vec r_3) 
\left( |\uparrow\uparrow\downarrow\rangle 
+ |\downarrow\uparrow\uparrow\rangle 
- 2 |\uparrow\downarrow\uparrow\rangle \right) &\deltaexp \cr
&\qquad\qquad\quad +
\psi^3_{11}(\vec r_1, \vec r_2, \vec r_3) 
\left( |\downarrow\uparrow\uparrow\rangle 
+ |\uparrow\downarrow\uparrow\rangle 
- 2 |\uparrow\uparrow\downarrow\rangle \right) \Bigr\} .
&\cr}$$
The notation $\psi^a_{\ell m}(\vec r_1, \vec r_2, \vec r_3)$ represents a
three-quark spatial wave function for which the $a^{\rm th}$ quark is in
the $|\ell, m\rangle$ orbitally excited state and the other two lie in the
ground state.  Note that this spatial wave function could also have been
written solely in terms of the relative coordinates, $\vec r_i-\vec r_j$.
The three positions of the quarks have been included to emphasize that this
function depends upon the center of mass coordinate.  We shall assume that
this dependence cancels when the terms are summed so that the final
baryonic state only depends on the quarks' relative
coordinates\footnote{\dag}{In \isgurkarl\ and \dalitz\ these coordinates
are usually written $\vec\rho = (\vec r_1 - \vec r_2)/\sqrt{2}$ and
$\vec\lambda = (\vec r_1 + \vec r_2 - 2 \vec r_3)/\sqrt{6}$.  $\vec r_i$ is
the position of the $i$th quark.}.  This assumption is certainly true for a
wide class of quark potential models including those of \isgurkarl
--\glozman .  This observation is important for understanding the $S_3$
transformation properties of the terms on each side of this equation.

\subsec{$S_3$.}

Since the permutation group $S_3$ plays an important role in this analysis
of the $L=1$ baryons, we review its basic properties at the same time
establishing our notation.  The group has three irreducible
representations: the trivial representation $S$ which corresponds to the
completely symmetric Young tableau $\stableau$, a one dimensional
representation $A$ that maps reflections to $-1$ and corresponds to the
tableau $\atableau$  and the two dimensional representation $M$ corresponding
to $\mtableau$.  The character table for the conjugacy classes of $S_3$---the
identity $e$, the three reflections $r$ and the two cyclic permutations
$c$---is
$$\beginpicture
\setcoordinatesystem units <0.5truein,0.25truein>
\setplotarea x from -1 to 5, y from -5.5 to 1.5
\linethickness=1.0pt
\putrule from -0.25 1.375 to 4.5 1.375
\linethickness=1.5pt
\putrule from -0.25 0 to 4.5 0
\linethickness=1.0pt
\putrule from -0.25 -4 to 4.5 -4
\putrule from 1.25 -4 to 1.25 0
\put {$e$} [c] at 2 0.625 
\put {$r$} [c] at 3 0.625 
\put {$c$} [c] at 4 0.625 
\put {$S$} [c] at 0.5 -1
\put {$A$} [c] at 0.5 -2
\put {$M$} [c] at 0.5 -3
\put {1} [c] at 2 -1
\put {1} [c] at 2 -2
\put {2} [c] at 2 -3
\put {1} [c] at 3 -1
\put {-1} [c] at 3 -2
\put {0} [c] at 3 -3
\put {1} [c] at 4 -1
\put {1} [c] at 4 -2
\put {-1} [c] at 4 -3
\put {Table 1.  Character Table for $S_3$.} [l] at -0.25 -5 
\endpicture$$
and from this table we can derive the following rules for the tensor
products of the irreducible representations:
$$\eqnn\multtable
\eqalignno{
S \otimes S = S \qquad S \otimes A = A \qquad S \otimes M &= M &\cr
A \otimes A = S \qquad A \otimes M &= M &\multtable\cr
M \otimes M &= S \oplus A \oplus M .&\cr}$$

Any three objects that may be permuted among each other form a three
dimensional, defining representation of $S_3$.  The positions of the three
quarks, $\{ \vec r_1, \vec r_2, \vec r_3 \}$, for example, form a ${\bf 3}$
of $S_3$.  This representation is not irreducible and can be separated into
the center of mass coordinate,
$$\vec R = {1\over\sqrt{3}} \left( \vec r_1 + \vec r_2 + \vec r_3 \right),$$
and a pair of coordinates for the internal motion, 
$$\pmatrix{ \vec r_+ \cr \vec r_-\cr} =
\pmatrix{
{1\over\sqrt{6}} \left( \vec r_1 + \vec r_2 - 2 \vec r_3 \right) \cr
{1\over\sqrt{2}} \left( \vec r_1 - \vec r_2 \right) \cr} .$$
This basis explicitly realizes the decomposition, ${\bf 3} = S \oplus M$.

The three-quark wave functions in equation \deltaexp\ transform as ${\bf 3}
= S \oplus M$ since they depend on all three positions.  The $L=1$ baryons,
however, transform as a two-dimensional $M$ representation, both under the
$S_3$ which corresponds to the spin-flavor group $SU(6)$ as well as the
$S_3$ referring to the spatial wave functions.  This representation for the
$L=1$ baryons, combined with those for the mass splitting operators
introduced in the next section, are the only information we require
of the spatial behavior of the quarks.  The $S_3$ group theory is
sufficiently restrictive to allow all the baryon matrix elements of mass
operators to be reduced to expressions involving a small number of
undetermined constants.

\newsec{Mass Operators.}

The definitions of the ground state baryon wave functions are essentially
the same for any non-relativistic quark model.  Only in the spatial wave
functions might one model differ from another---but what is important for
our approach is only the $S_3$ transformation properties, not the details
of the spatial dependence.  The lowest order differences among models
appear in the perturbations to the ground state functions.  The two
theories we compare here---the quark model and the chiral quark
model---have a similar set of operators which differ in the appearance of
flavor matrices in the interactions of the latter theory.

Our problem is to solve for the masses of the $L=1$ baryons in the
non-relativistic limit, $H\, |\Psi\rangle = E\, |\Psi\rangle$.  The
Hamiltonian, $H=H_0+V$, is assumed to be the sum of an $SU(6)$ symmetric
confining term $H_0$ which does not distinguish the masses of the 70-plet
and a perturbative potential $V$ that depends on the model being studied.
The first model that we study, both as an illustration of the method and a
benchmark against which to compare other models, is the constituent quark
model (\dgg\ and \isgurkarl ).  In this model the perturbative potential
arises from the first-order term in the expansion of a two-body Coulombic
interaction between pairs of quarks,\footnote{\dag}{Sometimes (\isgurkarl\ 
and \ic) the spin-spin and the quadrupole operator are grouped together and
called the hyperfine interaction, $V_{\rm hyp}=V_{ss}+V_q$.}
$$V= V_{ss} + V_{so} + V_q . \neqno\veqn$$
Traditionally\dgg , these interactions are of the form 
$$V_{ss} = \sum_{i<j} {16\pi\alpha_s\over 9} {1\over m_im_j}\, 
\vec s_i \cdot \vec s_j\, \delta(\vec r_{ij}), \neqno\vsst$$
a spin-spin interaction,
$$\eqnn\vsot \eqalignno{
V_{so} &= \sum_{i<j} {\alpha_s\over 3 r_{ij}^3}\, \biggl[
  {1\over m_i^2}\, (\vec r_{ij}\times \vec p_i)\cdot \vec s_i 
- {1\over m_j^2}\, (\vec r_{ij}\times \vec p_j)\cdot \vec s_j 
&\cr &\qquad\quad
+ {2\over m_im_j}\, (\vec r_{ij}\times \vec p_i)\cdot \vec s_j 
- {2\over m_im_j}\, (\vec r_{ij}\times \vec p_j)\cdot \vec s_i 
\biggr], &\vsot\cr}$$
a spin-orbit coupling, and 
$$V_q = \sum_{i<j} {2\alpha_s\over 3 r_{ij}^3} {1\over m_im_j}\, \biggl[
{3\over r_{ij}^2} (\vec r_{ij}\cdot \vec s_i) (\vec r_{ij}\cdot \vec s_j)
- (\vec s_i\cdot \vec s_j) \biggr] , \neqno\vqt$$
a quadrupole (or tensor) interaction.  Here, $\vec r_{ij}\equiv \vec r_i -
\vec r_j$ and $r_{ij}\equiv |\vec r_{ij}|$ where $\vec r_i$, $\vec p_i$,
$\vec s_i$, and $m_i$ are the position, momentum, spin, and mass of the
$i$th quark.  As with the quark wave functions, our treatment is
independent of the radial dependences of these potentials.  The only
important feature of these operators is their spin, orbital angular
momentum, and flavor structure.  Therefore, our analysis applies equally
well to any set of hyperfine and spin-orbit interactions of the form
$$\eqnn\vqm \eqalignno{
V_{ss} &= \sum_{i<j} f_0(r_{ij}) {1\over m_im_j}\, 
\vec s_i \cdot \vec s_j &\cr
V_{so} &= \sum_{i<j} f_1(r_{ij})\, \biggl[
  {1\over m_i^2}\, (\vec r_{ij}\times \vec p_i)\cdot \vec s_i 
- {1\over m_j^2}\, (\vec r_{ij}\times \vec p_j)\cdot \vec s_j 
&\cr &\qquad\qquad 
+ {2\over m_im_j}\, (\vec r_{ij}\times \vec p_i)\cdot \vec s_j 
- {2\over m_im_j}\, (\vec r_{ij}\times \vec p_j)\cdot \vec s_i 
\biggr] &\vqm \cr
V_q &= \sum_{i<j} f_2(r_{ij}) {1\over m_im_j}\, \biggl[
3 (\hat r_{ij}\cdot \vec s_i) (\hat r_{ij}\cdot \vec s_j)
- (\vec s_i\cdot \vec s_j) \biggr] , &\cr}$$
where the $f_a(r_{ij})$ are arbitrary functions of the distances between
the interacting quarks.  Note that we have included a factor of two in the
$1/m_im_j$ terms of the spin-orbit potential, $V_{so}$, to match the
non-relativistic limit for the potential.  We have retained this factor
here (and later in equation $(3.6)$) since when $SU(3)$ breaking effects,
such as a heavier constituent mass for the strange quark, are included in
$V_{so}$, the fits depend upon the choice of this factor.  In the limit
that the spin-dependent interactions are taken to be $SU(3)$-symmetric,
this dependence disappears and the factors of two can be replaced by an
arbitrary coefficient without affecting our results.

The second model we study is motivated by a recent proposal by Glozman and
his collaborators \glozman .  The idea is that as the typical momentum of a
quarks within a baryon is below the scale of chiral symmetry breaking, the
correct dynamical degrees of freedom are those of the constituent quarks
which couple to Goldstone boson fields of the broken symmetry group,
$SU(3)_L\times SU(3)_R \to SU(3)_V$.  This model, the chiral quark model,
modifies the low-energy Coulombic potential since the constituent
quark-Goldstone boson vertices carry additional $SU(3)$ flavor matrices,
$\lambda_i^a$:
$$\beginpicture
\setcoordinatesystem units <0.75truein,0.75truein>
\setplotarea x from -2 to 2, y from -1.5 to 1.25
\linethickness=1.0pt
\setlinear
\plot -1 -0.3  0 0  1 -0.3 /
\arrow <10pt> [0.2,0.67] from -1 -0.3 to -0.35 -0.105
\arrow <10pt> [0.2,0.67] from 0 0 to 0.6 -0.18
\linethickness=2.0pt
\setdashes
\plot 0 0  0 1 /
\put {$if\lambda^a$} [l] at 0.25 0.25
\put {Figure 1.  The constitent quark-Goldstone} [l] at -2.167 -0.75 
\put {boson vertex of the chiral quark model.} [l] at -2.167 -0.95 
\put {$\lambda^a$ is a Gell-Mann flavor matrix.} [l] at -2.167 -1.15 
\endpicture$$
This vertex produces perturbative potentials of the same form as in the
quark model except for the inclusion of a flavor factor, $\vec\lambda_i
\cdot \vec\lambda_j \equiv \sum_{a=1}^8 \lambda_i^a\, \lambda_j^a$:
$$\eqnn\vcqm \eqalignno{
V_{ss} &= \sum_{i<j} g_0(r_{ij}) 
\left\{ {1\over m_im_j}, \vec\lambda_i\cdot\vec\lambda_j \right\} \, 
\vec s_i \cdot \vec s_j &\cr
V_{so} &= \sum_{i<j} g_1(r_{ij})\, \biggl[
  \left\{ {1\over m_i^2}, \vec\lambda_i\cdot\vec\lambda_j \right\} \,  
   (\vec r_{ij}\times \vec p_i)\cdot \vec s_i 
- \left\{ {1\over m_j^2}, \vec\lambda_i\cdot\vec\lambda_j \right\} \, 
   (\vec r_{ij}\times \vec p_j)\cdot \vec s_j
&\cr &\qquad\qquad 
+ \left\{ {2\over m_im_j}, \vec\lambda_i\cdot\vec\lambda_j \right\} \, 
   (\vec r_{ij}\times \vec p_i)\cdot \vec s_j 
- \left\{ {2\over m_im_j}, \vec\lambda_i\cdot\vec\lambda_j \right\} \, 
   (\vec r_{ij}\times \vec p_j)\cdot \vec s_i 
\biggr] &\vcqm\cr
V_q &= \sum_{i<j} g_2(r_{ij}) 
\left\{ {1\over m_im_j}, \vec\lambda_i\cdot\vec\lambda_j \right\} \, 
\biggl[ 3 (\hat r_{ij}\cdot \vec s_i) (\hat r_{ij}\cdot \vec s_j)
- (\vec s_i\cdot \vec s_j) \biggr] , &\cr}$$
where we have written the arbitrary functions as $g_a(r_{ij})$ to emphasize
that they need not be the same as those in the quark model.  The
anticommutators ensure that the operators are Hermitian.

\subsec{$S_3$ Transformation Properties of the Mass Operators.}

The fact that the states and the potentials in equations \vqm\  and \vcqm\ 
transform as definite representations of $S_3$ allows us to constrain
greatly the number of unknown parameters in the theory.  We therefore first
present a method for counting the number of independent constants before
writing them in a more concrete form:  as matrix elements of specific
operators between three-quark states.  Both pieces of the hyperfine
interaction, the spin-spin and the quadrupole operators, transform as
three dimensional representations of $S_3$; both are manifestly symmetric
under exchanging the interacting quarks.  In terms of the irreducible
representations of $S_3$, we saw that the {\bf 3} could be decomposed as
$${\bf 3}=S\oplus M, \neqno\three$$
that is, both the spin-spin operator and the quadrupole operator have a
piece transforming as the trivial representation and a piece transforming
as the two-dimensional representation.  The matrix elements $\langle {\bf
70}\, |\, V_{ss}\, |\, {\bf 70}\rangle$ or $\langle {\bf 70}\, |\, V_q\,
|\, {\bf 70}\rangle$, which produce the perturbations to the baryon mass
spectrum, contain the following tensor product of $S_3$
representations:
$$M \otimes (S\oplus M) \otimes M =
\underbrace{S \oplus S}_{\rm trivial} 
\oplus A \oplus A \oplus M \oplus M \oplus M \oplus M .
\neqno\hyptensor $$
The trivial representation appears twice in the matrix element.  From the
Wigner-Eckart theorem, we can conclude that the matrix elements of $V_{ss}$
and $V_q$ are each completely determined up to two unknown constants.  In
specific models, these constants correspond physically to spatial
integrals.  

The spin-orbit term, $V_{so}$, is slightly more complicated.  It transforms
as a six dimensional fundamental representation of $S_3$.  The
decomposition of the {\bf 6} into its irreducible components is 
$${\bf 6} = S \oplus A \oplus M \oplus M .\neqno\six$$
Counting the number of unknown constants, we learn that the matrix elements
of $V_{so}$ between {\bf 70} states,
$$\langle {\bf 70}\, |\, V_{so}\, |\, {\bf 70}\rangle \to 
M \otimes {\bf 6} \otimes M = \underbrace{S \oplus S \oplus S \oplus
S}_{\rm trivial} \oplus A \oplus A \oplus A \oplus A \oplus
\overbrace{M \oplus \cdots \oplus M}^{\rm 8\ copies} ,
\neqno\sotensor$$
depend upon four undetermined constants since the trivial representation
appears four times.  

It is helpful to have a specific form for these independent integrals which
can be calculated for a particular model for the quark wave functions.  The
spin-orbit operator being the most complicated, we begin with it.  It can
be written in the form\footnote{\dag}{When the masses of the quarks are
equal in $V_{so}$, the factors of two for the $1/m_im_j$ terms can be
replaced by an arbitrary coefficient without altering the results that
follow.  The expression in the chiral quark model is analogous.}
$$V_{so} = \sum_{i<j}\, \biggl[
  {1\over m_i^2} \vec {\cal L}_{ij}\cdot \vec s_i 
+ {2\over m_im_j}\vec {\cal L}_{ji}\cdot \vec s_i 
+ {2\over m_im_j}\vec {\cal L}_{ij}\cdot \vec s_j 
+ {1\over m_j^2} \vec {\cal L}_{ji}\cdot \vec s_j 
\biggr] \neqno\vsogen$$
where 
$$\vec{\cal L}_{ij} \equiv f_1(r_{ij})\, \left[ 
\vec L_i - (\vec r_j \times \vec p_i) \right] \neqno\lij$$
$\vec L_i$ being the orbital angular momentum of the $i$th quark.  In the
non-relativistic limit, the spin and the flavor structures are completely
calculable---what is relevant for the $S_3$ group theory is the spatially
dependent operator $\vec{\cal L}_{ij}$.  In the matrix element $\langle
{\bf 70}\, |\, V_{so}\, |\, {\bf 70}\rangle$ appear sums of matrix elements 
of this operator between $L=1$ three-quark states,
$$\langle \psi^a_{1m}(\vec r_1, \vec r_2, \vec r_3)\, |\, \vec{\cal L}_{12}\,
|\, \psi^b_{1m'}(\vec r_1, \vec r_2, \vec r_3) \rangle =
{\cal M}^{ab}\, \langle 1,m\, |\, \vec L\, |\, 1,m' \rangle .
\neqno\mabdef$$
where $\vec L$ is the total orbital angular momentum of the three-quark
state.  The matrix elements for other choices of $i$ and $j$ for $\vec
{\cal L}_{ij}$ are similarly defined but are simply a permutation of the
entries of the elements of ${\cal M}^{ab}$.  The matrix ${\cal M}^{ab}$ is
a $3\times 3$ Hermitian matrix of spatial integrals differing only as to
which of initial or final the quarks is excited.  We know from the
Wigner-Eckart theorem that the matrix elements of the 70-plet baryons
cannot depend separately upon all of the elements of the matrix ${\cal M}$,
but only upon four linear combinations; these combinations are
$$\eqalign{
{\cal SO}_1 &\equiv 
{\cal M}^{11} + {\cal M}^{22} - {\cal M}^{12} - {\cal M}^{21}\cr
{\cal SO}_2 &\equiv 
{\cal M}^{11} - {\cal M}^{22} - {\cal M}^{13} + {\cal M}^{23}
- {\cal M}^{31} + {\cal M}^{32}\cr
{\cal SO}_3 &\equiv 
{\cal M}^{11} + {\cal M}^{22} + 4{\cal M}^{33} + {\cal M}^{12} + {\cal M}^{21}
- 2{\cal M}^{13} - 2{\cal M}^{23} - 2{\cal M}^{31} - 2{\cal M}^{32}\cr
{\cal SO}_4 &\equiv 
({\cal M}^{12} - {\cal M}^{21}) - 
({\cal M}^{13} - {\cal M}^{31}) + ({\cal M}^{23} - {\cal M}^{32}) .\cr}$$
The matrix ${\cal M}$ being Hermitian, ${\cal SO}_4$ must be purely
imaginary or zero.  

Time reversal provides an additional constraint which imposes 
${\cal SO}_4\equiv 0$.  Under time reversal, we assume that the three quark
wave functions transform in the usual way:  that
$$\Theta\, \psi^a_{\ell m}(\vec r_1, \vec r_2, \vec r_3) = 
(-1)^{\ell+m}\, \psi^a_{\ell m}(\vec r_1, \vec r_2, \vec r_3) ,
\neqno\timerev$$
where $\Theta$ represents the time reversal operator, and that $\vec{\cal
L}_{ij}$ transforms as $\Theta\vec{\cal L}_{ij}\Theta^{-1} = - \vec{\cal
L}_{ij}$.  It follows that ${\cal M}^{ab}={\cal M}^{ba}$ is a symmetric
matrix.  Since ${\cal M}^{ab}=({\cal M}^{ba})^*$, ${\cal M}^{ab}$ is
therefore a real symmetric matrix and the linear combination ${\cal SO}_4$
must vanish.  Thus, for both the quark model and the chiral quark model,
the spin-orbit matrix elements of the barons are completely determined by
only three constants ${\cal SO}_1$, ${\cal SO}_2$ and ${\cal SO}_3$, which
can be calculated in a specific model.

We describe the hyperfine interactions more briefly.  As we know in advance
from the $S_3$ group theory that there are only two independent constants
in either case, we shall define fewer three quark matrix elements than was
done for the spin-orbit operator.  Let us define the spatial integrals
$$\eqnn\ssdef \eqalignno{
\langle \psi^1_{1m}(\vec r_1, \vec r_2, \vec r_3)\, |\, f_0(r_{12})\,
|\, \psi^1_{1m'}(\vec r_1, \vec r_2, \vec r_3) \rangle 
&= {\cal A}_1\, \delta_{mm'} &\cr
\langle \psi^1_{1m}(\vec r_1, \vec r_2, \vec r_3)\, |\, f_0(r_{12})\,
|\, \psi^2_{1m'}(\vec r_1, \vec r_2, \vec r_3) \rangle 
&= {\cal A}_2\, \delta_{mm'} &\ssdef\cr
\langle \psi^3_{1m}(\vec r_1, \vec r_2, \vec r_3)\, |\, f_0(r_{12})\,
|\, \psi^3_{1m'}(\vec r_1, \vec r_2, \vec r_3) \rangle 
&= {\cal A}_3\, \delta_{mm'} &\cr}$$
for the spin-spin operator.  Then only the following linear combinations
appear in the matrix elements for the baryons:
$$\eqnn\ddef \eqalignno{
{\cal D}_+ &= {\cal A}_1 + {\cal A}_2 + 2{\cal A}_3 &\cr
{\cal D}_- &= {\cal A}_1 - {\cal A}_2 . &\ddef \cr}$$
A sufficient basis of spatial integrals for writing all of the baryon
matrix elements of the quadrupole operator is provided simply by
$$\eqnn\bdef \eqalignno{
\langle \psi^1_{1m}(\vec r_1, \vec r_2, \vec r_3)\, |\, Q^{\alpha\beta}\, 
|\, \psi^1_{1m'}(\vec r_1, \vec r_2, \vec r_3) \rangle
&= {\cal Q}_1\, \langle 1,m\, |\, 
L^\alpha L^\beta - {\textstyle{1\over 3}}
\delta^{\alpha\beta} L^2\, |\, 1,m' \rangle &\cr
\langle \psi^1_{1m}(\vec r_1, \vec r_2, \vec r_3)\, |\, Q^{\alpha\beta}\, 
|\, \psi^2_{1m'}(\vec r_1, \vec r_2, \vec r_3) \rangle
&= {\cal Q}_2\, \langle 1,m\, |\, 
L^\alpha L^\beta - {\textstyle{1\over 3}}
\delta^{\alpha\beta} L^2\, |\, 1,m' \rangle .
&\bdef \cr}$$
where
$$Q^{\alpha\beta} \equiv f_2(r_{12})\left( 
3\, \hat r_{12}^\alpha \hat r_{12}^\beta - \delta^{\alpha\beta} 
\right) \neqno\qab $$
Neither of these cases is further constrained by time-reversal symmetry
other than to say that the above constants are all real.

It is now possible to express the masses of the $L=1$ baryons in terms of
the seven unknown spatial integrals (${\cal D}_\pm$, ${\cal SO}_{1,2,3}$,
${\cal Q}_{1,2}$) and the common 70-plet zero-order mass $\langle {\bf
70}\, |\, H_0\, |\, {\bf 70}\rangle\equiv M_0$.  We further shall
explicitly break $SU(3)$, while keeping isospin symmetry, by giving the
strange quark a larger mass, $m_s> m_u=m_d$.  These masses correspond to
the constituent masses so that the strange-up mass difference can be
assumed to be small,
$$\delta m \equiv {m_s-m_u\over m_u} \ll 1. \neqno\dm $$
Since $SU(3)$ is only weakly broken, we shall work only to first order in
$\delta m \approx 0.27$--$0.29$.  The mass factors appear explicitly in the
$1/m_i$ coefficients of the perturbative potentials.

We should here pause to remark on the power of the $S_3$ argument.  The
group theory allows us to calculate the perturbations to a model with a
specific spin, orbital angular momentum and flavor structure but with other
details left arbitrary.  This feature allows us to test the plausibility of
a model's ability to explain the observed $L=1$ baryon spectrum by
adjusting the independent constants to fit these masses.  If the model
fails to fit the data to a reasonable confidence level, then regardless of
the dynamical model for the quark wave functions used, it will still fail
adequately to generate the observed $L=1$ mass splittings.  The converse,
however, is not true.  Even should a model fit the data well, realistic
choices for the quark wave functions may not achieve the best fit
attainable in the full parameter space.

As an example, the mass perturbation to the $|\Delta; {\textstyle{1\over 2}}; 
J={\textstyle{3\over 2}} \rangle$ states in the quark model is
$$\Delta m_{\Delta,3/2} = M_0 + {1\over 2} {\cal D}_1 - {9\over 2} {\cal D}_2 
+ {\cal SO}_2 + {\cal SO}_3 \neqno\dmexp$$
while for the chiral quark model, the perturbation becomes
$$\Delta m_{\Delta,3/2} = M_0 + {2\over 3}{\cal D}_1 - 6 {\cal D}_2 
+ {4\over 3} {\cal SO}_2 + {4\over 3} {\cal SO}_3 .\neqno\dmexpf$$
The calculation of these matrix elements, as well as those for the rest of
the baryons was accomplished with the Maple symbolic manipulation program.
The mass splitting operators will in general mix baryons which have
equivalent total angular momentum, isospin and strangeness.

\newsec{The Comparison with Experiment.}

Among the baryons observed to date, eighteen have been reliably identified
with the $L=1$ baryons \ref\pdg{Particle Data Group, ``Review of Particle
Physics,'' Eur.~Phys.~J.~{\bf C3} (1998) 1.}.  Our program then is to
obtain the best possible fit with the nine unknown quantities---the seven
spatial integrals for the interactions, a parameter for the $SU(3)$
breaking, and zero-order baryon mass---to the masses of these eighteen
baryons.  Fits were made for each of the two models.  The actual fitting
routine applied a Levenberg-Marquardt algorithm \ref\nr{W.~H.~Press,
S.~A.~Teukolsky, W.~T.~Vetterling, and B.~P.~Flannery, {\it Numerical
Recipes in Fortran\/}, Second Edition, Cambridge University Press, 1992,
\S 15.5.}  which chose its initial conditions randomly within this 
nine-dimensional parameter space.  For those baryons within an incompletely
measured set with the same $J$, $I$, and strangeness, specifically the
$\Sigma_{J=3/2}$, the $\Sigma_{J=1/2}$, the $\Lambda_{J=3/2}$, and the
$\Xi_{J=3/2}$, all the possible assignments of elements of these sets to
the measured masses were sampled; those which produced the best fit to the
mass spectrum are displayed in this section.  The values for the parameters
for the two pictures of the low energy physics within a baryon are shown in
Table 2, with the $\chi^2$ for each of the fits:
$$\beginpicture
\setcoordinatesystem units <0.5truein,0.25truein>
\setplotarea x from -1 to 5, y from -15 to 2.5
\linethickness=1.0pt
\putrule from -1 2.25 to 5 2.25
\linethickness=1.5pt
\putrule from -1 0 to 5 0
\linethickness=1.0pt
\putrule from -1 -11 to 5 -11
\putrule from 1 -11 to 1 0
\put {Perturbation} [c] at 0 1.5 
\put {(MeV)$^*$} [c] at 0 0.625 
\put {Quark} [c] at 2 1.5
\put {Model} [c] at 2 0.625
\put {Chiral} [c] at 4 1.5
\put {Quark Model} [c] at 4 0.625
\put {${\cal D}_+$} [c] at 0 -1
\put {${\cal D}_-$} [c] at 0 -2
\put {${\cal SO}_1$} [c] at 0 -3
\put {${\cal SO}_2$} [c] at 0 -4
\put {${\cal SO}_3$} [c] at 0 -5
\put {${\cal Q}_1$} [c] at 0 -6
\put {${\cal Q}_2$} [c] at 0 -7
\put {$\delta m$} [c] at 0 -8
\put {$M_0$} [c] at 0 -9
\put {$\chi^2$} [c] at 0 -10
\put {196.9} [c] at 2 -1
\put {-19.50} [c] at 2 -2
\put {30.85} [c] at 2 -3
\put {47.99} [c] at 2 -4
\put {-90.50} [c] at 2 -5
\put {12.63} [c] at 2 -6
\put {7.762} [c] at 2 -7
\put {0.269} [c] at 2 -8
\put {1613} [c] at 2 -9
\put {123.6} [c] at 2 -10
\put {-87.78} [c] at 4 -1
\put {-52.11} [c] at 4 -2
\put {-1.106} [c] at 4 -3
\put {15.93} [c] at 4 -4
\put {17.40} [c] at 4 -5
\put {-14.10} [c] at 4 -6
\put {-15.38} [c] at 4 -7
\put {0.286} [c] at 4 -8
\put {1477} [c] at 4 -9
\put {24.76} [c] at 4 -10
\put {Table 2.  Best-fit values of the $S_3$ constants for} [l] at -1 -12 
\put {the quark model and chiral quark model.  $^*$All} [l] at -1 -12.8
\put {of the parameters are in units of MeV except} [l] at -1 -13.6
\put {$\delta m$ which is dimensionless.} [l] at -1 -14.4
\endpicture$$
These values produce the mass spectra displayed in figure 2 for the quark
model and in figure 3 for the chiral quark model.  In these figures, the
baryons used to fit the experimentally observed masses are shown in
unbroken lines while the remaining baryons masses, shown in dashed lines,
represent predictions.  The composition of the baryons in terms of
spin-flavor multiplets is summarized in tables 3 and 4.

The eighteen baryons chosen for the fits are those listed in the Baryon
Summary Table of \pdg ; we should mention that the existence two other
$L=1$ states, the $\Sigma_{J=1/2}(1620)$ and the $\Sigma_{J=3/2}(1580)$,
has been fairly well established.  As we shall see, the quark model is
unable to fit the baryon spectrum with a satisfactory $\chi^2$ even with
the omission of these two baryons.  But as the chiral quark model
successfully fit the baryons listed in the Baryon Summary Table, we
have included these two in table 4.  Although they were not included in the
fitting routine, each baryon has a `predicted' state within 20--40 MeV so that
we do not expect that our conclusions, nor the $\chi^2$ of the fit, would
alter greatly had they been included in the fitting procedure.

\subsec{The Quark Model.}

The standard quark model fares rather poorly with a best $\chi^2$ value of
123.6 for only eighteen fit parameters.  The masses of the $\Lambda(1405)$
[$J=\half$] and the $\Lambda(1520)$ [$J=\thalf$] baryons have been been
measured to within $\pm 4$ MeV and $\pm 1$ MeV respectively and tend to
drive the fit parameters to produce a precise fit for these states---at the
expense of others.  An accurate fit of the $\Lambda$'s tends to produce a
poor fit for the decuplet states, most glaringly, giving a predicted mass of
1884 MeV for the $\Delta(1620)$.  More generally, the quark model predicts
higher masses for decuplet states with lower total angular momentum in
contrast with the general trend for the $\ell=1$ baryons, in particular the
observed reversed ordering of the $J=\half$ and $J=\thalf$ $\Delta$ masses.
The original study of the 70-plet baryons by Isgur and Karl \isgurkarl\
succeeded in obtaining a better fit for the decuplet states but only at the
expense of a predicted mass of 1490 MeV for the $\Lambda(1405)$ and a
consequently poorer $\chi^2$.  It seems difficult for the constituent quark
model to account for {\it both\/} the lightness of the $\Lambda$ states and
the decuplet mass spectrum.  This failure is often described in terms of
the size of the spin-orbit coupling:  a weak coupling is needed to fit the
majority of the baryons but a strong coupling is required to generate the
observed $\Lambda(1520)$--$\Lambda(1405)$ mass splitting 
\ref\reinders{L.~J.~Reinders, ``Spin-Orbit Forces in the Baryon Spectrum,'' 
in {\it Baryon 1980\/}, proceedings of the IVth International Conference on
Baryon Resonances, Toronto, 1980, edited by N.~Isgur (University of Toronto
Press, Toronto, 1981) p.~203.}.

\subsec{The Chiral Quark Model.}
The chiral quark model is able to reconcile successfully these two features
and produce an acceptable fit for the detected baryon spectrum: a $\chi^2$
of only 24.76.  Its worst failure among the observed baryons is that the model
does not generate a sufficient splitting in the $J=\thalf$ $N$ states, only
about 20 MeV compared with an experimental splitting of almost 200 MeV.  At
present, the large experimental error in the $N(1700)$ allows a ``good''
fit to be achieved but it may be difficult to accommodate the actual
splitting as more precise data are obtained.

As mentioned, the quark model analysis of Isgur and Karl \isgurkarl\ found
that, aside from large $\Lambda(1520)$--$\Lambda(1405)$ mass splitting, the
splitting among the baryon multiplets required an extremely small
spin-orbit contribution to the masses.  The difficultly in justifying this
small spin-orbit coupling has been called the ``spin-orbit puzzle.''  In the
chiral quark model, while the spin-spin interactions dominate with a
strength roughly five times that of the other terms, the spin-orbit
interactions are comparable to the quadrupole interactions, with only two
of the three independent integrals responsible for essentially all of the
spin-orbit contribution.  Thus the spin-orbit puzzle does not seem to occur in
the chiral quark model.

\subsec{The $\Lambda(1405)$.}

The low mass of the $\Lambda(1405)$ has marked it as something of a
conundrum among the $L=1$ baryons.  At one extreme, it would seem
reasonable to regard it as a $\bar KN$ bound state since its mass lies 30
MeV below the $\bar KN$ threshold.  Alternatively, the non-relativistic quark
model treats the $\Lambda(1405)$ as an ordinary $L=1$ baryon composed
of some mixture of $SU(3)$ singlet and octet states.  Traditionally, the quark
model \isgurkarl\ predicts that the $\Lambda(1405)$ mainly is composed of
the singlet state with a small admixture of the octet states and such a
behavior is seen in the best-fit results for the quark model (gluon exchange):
$$\Lambda(1405)_{qm} = 
- 0.9998\, | ^21;\half;\half\rangle
+ 0.0082\, | ^28;\half;\half\rangle
+ 0.0178\, | ^48;\thalf;\half\rangle . \neqno\lamqm$$
The chiral quark model (Goldstone boson exchange) gives a similar result
for the composition of the $\Lambda(1405)$ except that the spin-$\half$
contribution is slightly enhanced:
$$\Lambda(1405)_{\chi qm} = 
- 0.9775\, | ^21;\half;\half\rangle
- 0.2071\, | ^28;\half;\half\rangle
- 0.0395\, | ^48;\thalf;\half\rangle . \neqno\lamcqm$$
Since the chiral quark model succeeds in fitting the observed baryon masses
well, it is instructive to probe the model further by comparing the
consequences of this predicted composition with some of the other
phenomenological properties of this baryon.

Nathan Isgur \ref\isgur{N.~Isgur, in {\it Baryons '95\/}, proceedings of
the 7th International Conference on the Structure of Baryons, edited by
B.~F.~Gibson, {\it et.\ al.\/}\ (World Scientific, 1996).}\  has recently
proposed that the $\Lambda(1405)$ can be studied in heavy quark effective
theory limit.  In this picture, the $\Lambda(1405)$ is a $uds$ quark bound
state where the strange quark mass is taken to be heavy compared to the up
and the down quark masses.  Singling out the $s$ quark breaks the $SU(3)$
flavor symmetry and its spin and orbital angular momentum completely
determine that of the $\Lambda(1405)$.  The $ud$ quarks form an inert
$S=0$, $L=0$ pair.  Such a state no longer can be described in terms of
pure $SU(3)$ states; however it does contain equal amounts of the singlet
and spin $\half$ octet states.  Such a composition contrasts with that
emerging in either the quark model or the chiral quark model.  In both
cases, the $\Lambda(1405)$ remains essentially a singlet state although the
chiral quark model does match the heavy quark theory's predictions
marginally better:
$$\langle \Lambda(1405)_{\chi qm}\, |\, \Lambda(1405)_{HQET} \rangle = 
0.6079 \neqno\overlapqm $$
compared to an overlap of 
$$\langle \Lambda(1405)_{qm}\, |\, \Lambda(1405)_{HQET} \rangle = 
0.5089 \neqno\overlapcqm$$
for the quark model.

\subsec{Mixing Angles from $L=1$ Decays.}

The decays of the {\bf 70} states into {\bf 56} states provide another
estimate of the observed baryons' compositions in terms of $SU(3)\times
SU(2)$ eigenstates.  While fits to the decay amplitudes have been performed
for the quark model \ref\koniuk{R.~Konuik and N.~Isgur, ``Baryon decays in a
quark model with chromodynamics,'' Phys.\ Rev.\ {\bf D21} (1980) 1868.},
they do not include estimates for the mixing angles; however, the mixing
angles have been extracted for the $SU(6)_W$ model \ref\hey{A.~J.~G.~Hey,
P.~J.~Litchfield and R.~J.~Cashmore, ``$SU(6)_W$ and Decays of Baryon
Resonances,'' Nucl.\ Phys.\ {\bf B95} (1975) 516.}, which has the same
algebraic structure for the decays as the standard quark model $SU(6)$.  A
comparison of the compositions of the states fit to {\bf 70} decays \hey\
with those of the two models fit here to the mass spectrum are shown in
Table 8.  In this table we have only included the states from mixed
$J$-multiplets that have been completely observed---the $N$ states and the
three $\Lambda_{J=1/2}$ states.
$$\beginpicture
\setcoordinatesystem units <3.0truein,0.25truein>
\setplotarea x from -1 to 1, y from -9 to 1.5
\linethickness=1.0pt
\putrule from -1 1.4 to 1 1.4
\linethickness=1.5pt
\putrule from -1 0 to 1 0
\linethickness=1.0pt
\putrule from -1 -6 to 1 -6
\putrule from 0 -6 to 0 0
\put {Quark Model} [c] at -0.5 0.625
\put {Chiral Quark Model} [c] at 0.5 0.625
\put {$_{\rm decays}\langle N_{J=1/2}\, |\, N_{J=1/2}\rangle_{qm} 
= -.98$} [c] at -0.5 -1
\put {$_{\rm decays}\langle N_{J=3/2}\, |\, N_{J=3/2}\rangle_{qm} 
= -.87$} [c] at -0.5 -2
\put {$_{\rm decays}\langle \Lambda(1405)\, |\, \Lambda(1405)\rangle_{qm} 
= -.84$} [c] at -0.5 -3
\put {$_{\rm decays}\langle \Lambda(1670)\, |\, \Lambda(1670)\rangle_{qm} 
= -.63$} [c] at -0.5 -4
\put {$_{\rm decays}\langle \Lambda(1800)\, |\, \Lambda(1800)\rangle_{qm} 
= -.70$} [c] at -0.5 -5
\put {$_{\rm decays}\langle N_{J=1/2}\, |\, N_{J=1/2}\rangle_{\chi qm} 
= -1.00$} [c] at 0.5 -1
\put {$_{\rm decays}\langle N_{J=3/2}\, |\, N_{J=3/2}\rangle_{\chi qm} 
= -.94$} [c] at 0.5 -2
\put {$_{\rm decays}\langle \Lambda(1405)\, |\, \Lambda(1405)\rangle_{\chi qm}
= -.94$} [c] at 0.5 -3
\put {$_{\rm decays}\langle \Lambda(1670)\, |\, \Lambda(1670)\rangle_{\chi qm}
= -.42$} [c] at 0.5 -4
\put {$_{\rm decays}\langle \Lambda(1800)\, |\, \Lambda(1800)\rangle_{\chi qm}
= -.48$} [c] at 0.5 -5
\put {Table 8.  A comparison of the compositions of the baryons obtained 
from the $SU(6)_W$} [l] at -1 -7
\put {model fit to ${\bf 70}\to {\bf 56} +\cdots$ decays \hey\  with those 
obtained in our fits for both the quark } [l] at -1 -7.8
\put {model ($qm$) and the chiral quark model 
($\chi qm$).} [l] at -1 -8.6
\endpicture$$
Both models agree extremely well with the decay estimates for the $N$ state
compositions, but they begin to disagree for the $\Lambda$
states.  The decays of the $L=1$ baryons suggest that the $\Lambda(1405)$
and the $\Lambda(1520)$ are principally singlet states, which is in accord
with our fits.  However, the $\Lambda(1800)$ in the chiral quark model is
predominantly an $S=\thalf$ octet state whereas the fits to the decay
amplitudes suggest it is principally an $S=\half$ octet state.  While not
conclusive, these disagreements suggest it may be a challenge for the chiral
quark model to fit simultaneously the mass spectrum and
the observed decay amplitudes.

\subsec{$SU(3)$ Symmetric Perturbations.}

In evaluating the matrix elements of the mass operators of equations \vqm\
and \vcqm, we explicitly broke flavor symmetry by giving the strange quark
a heavier mass.  We shall now examine what happens to the fits when $SU(3)$
is preserved in these spin operators.  The rationale for taking this limit
is that if both the spin splitting and the flavor breaking effects are
small, terms that simulaneously break $SU(3)$ and $SU(2)$ can be regarded
as higher order effects.

The results for the fits to the $L=1$ baryon spectrum due to an explicit
$SU(3)$ breaking term plus flavor-symmetric versions of the operators in
equations \vqm\ and \vcqm\ are shown in figures 4 and 5.  The mixing angles
are included in tables 6 and 7 while table 5 displays the best fit values
of the $S_3$ constants in units of MeV (except for the dimensionless
$SU(3)$ breaking parameter).  Surprisingly, the $\chi^2$ improves for the
quark model fit, from 123 to 79, although the general pattern remains as
before.  Some of the assignments of the states being fit, among the
$\Sigma_{J=1/2}$, $\Sigma_{J=3/2}$ and $\Lambda_{J=3/2}$ states, have
changed.  The ordering of the multiplets of decuplet remains unaltered.
One feature that figure 4 does not convey is that many other arrangments of
the baryons in the incompletely observed multiplets also lead to a better
fit than that of figure 2, the best fit obtained for the
$SU(3)$-breaking spin operators.

The value of $\chi^2$ for the chiral quark model predictably worsened when
$SU(3)$ was imposed on the spin-splitting operators.  The pattern of masses
otherwise did not change significantly.  Comparing the the masses in the
$SU(3)$ symmetric and the $SU(3)$-broken limits (tables 7 and 4) provides
an estimate for the theoretical errors associated with our fits---most of
the fit baryon masses agreed to within 10--20 MeV.  Most of the sizes of
the fit parameters did not differ much between the two fits with the
exception of the quadrupole interaction which is substantially smaller in
the $SU(3)$ symmetric limit.

\newsec{Conclusion.}

The $S_3$ permutation group provides a new tool for the study of the
physics within baryons.  In addition to freeing us from a specific choice
for the quark-quark potential, this approach allows a comparison of
theories differing in the flavor structure of their interactions.  When
applied to the traditional quark model and the more recent chiral quark
model, our approach places the two theories on an equal footing with a
one-to-one mapping of the unknown $S_3$ parameters between the two
theories.  The results of this comparison were somewhat surprising---the
chiral quark model shows a clearly better fit with the observed $L=1$
baryon spectrum.

Since the chiral quark model provided a good fit and seems to be able to
avoid the spin-orbit problem, we should mention some of the challenges that
it still faces.  As stated earlier, the fitting routine ranged over the
entire available parameter space and it remains to show that the best-fit
set of parameters can be realized by a physical potential for the
quark-quark interactions.  It would also be interesting to see whether this
superiority over the traditional quark model fit persists when we attempt
to fit simultaneously the $L=0$, the negative and positive parity $L=1$,
and lowest excited states of the $N=2$ band.  This program was carried out by
Capstick and Isgur \ic\ for a relativized quark model with harmonic
oscillator wave functions.  Finally, if the model is to provide a
believable explanation of the low-energy physics within a baryon it must
not only describe the mass spectrum, but also accommodate the excited state
decays.

\bigskip
\centerline{\bf Acknowledgements.}
We would like to thank Nathan Isgur and Gabriel Karl for their suggestions
of useful references on the early work about the quark model and the baryon
spectrum.

\footatend\vfill\supereject\immediate\closeout\rfile\writestoppt
\baselineskip=14pt\centerline{{\bf References}}\bigskip{\frenchspacing%
\parindent=20pt\escapechar=` \input refs.tmp\vfill\eject}\nonfrenchspacing
\vfill\eject

$$\beginpicture
\setcoordinatesystem units <0.225truein,0.5truein>
\setplotarea x from -1 to 25, y from 0 to 12
\putrule from 0 12  to 0 3
\putrule from 0 3  to 25 3
\putrule from -0.1 3 to 0 3
\putrule from -0.1 4 to 0 4
\putrule from -0.1 5 to 0 5
\putrule from -0.1 6 to 0 6
\putrule from -0.1 7 to 0 7
\putrule from -0.1 8 to 0 8
\putrule from -0.1 9 to 0 9
\putrule from -0.1 10 to 0 10
\putrule from -0.1 11 to 0 11
\putrule from -0.1 12 to 0 12
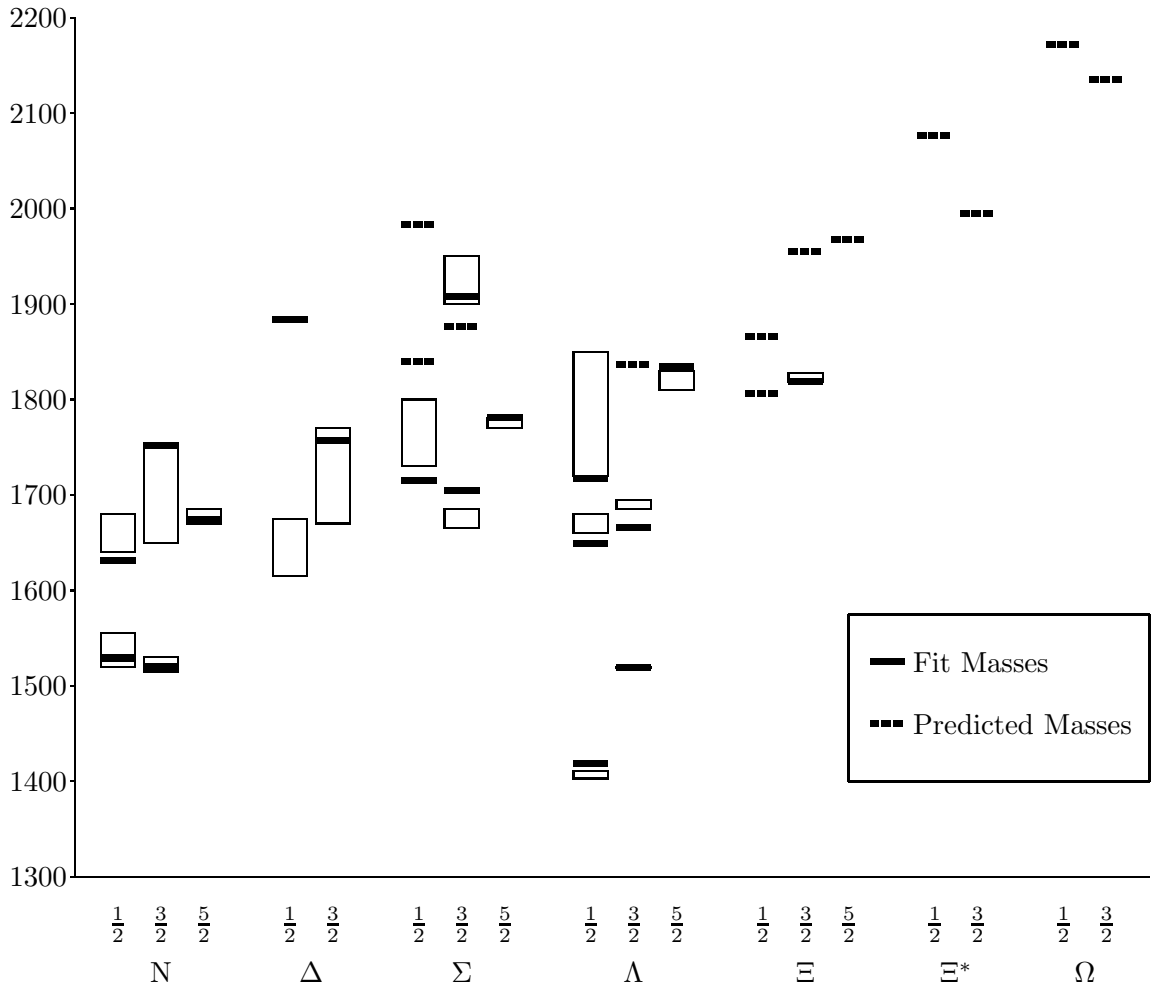
\put {Figure 2.  Masses of the $L=1$ Baryons in the Quark Model;
$\chi^2 = 123.6432$, $N_{fit} = 18$} [l] at -1 1.25
%
%
\put {1300} at -0.85 3 
\put {1400} at -0.85 4 
\put {1500} at -0.85 5 
\put {1600} at -0.85 6 
\put {1700} at -0.85 7 
\put {1800} at -0.85 8 
\put {1900} at -0.85 9 
\put {2000} at -0.85 10 
\put {2100} at -0.85 11 
\put {2200} at -0.85 12 
%
%
%
\put {N} at 2 2.0
\put {${\textstyle {1\over 2}}$} at 1 2.5
\put {${\textstyle {3\over 2}}$} at 2 2.5
\put {${\textstyle {5\over 2}}$} at 3 2.5
\put {$\Delta$} at 5.5 2.0
\put {${\textstyle {1\over 2}}$} at 5 2.5
\put {${\textstyle {3\over 2}}$} at 6 2.5
\put {$\Sigma$} at 9 2.0
\put {${\textstyle {1\over 2}}$} at 8 2.5
\put {${\textstyle {3\over 2}}$} at 9 2.5
\put {${\textstyle {5\over 2}}$} at 10 2.5
\put {$\Lambda$} at 13 2.0
\put {${\textstyle {1\over 2}}$} at 12 2.5
\put {${\textstyle {3\over 2}}$} at 13 2.5
\put {${\textstyle {5\over 2}}$} at 14 2.5
\put {$\Xi$} at 17 2.0
\put {${\textstyle {1\over 2}}$} at 16 2.5
\put {${\textstyle {3\over 2}}$} at 17 2.5
\put {${\textstyle {5\over 2}}$} at 18 2.5
\put {$\Xi^*$} at 20.5 2.0
\put {${\textstyle {1\over 2}}$} at 20 2.5
\put {${\textstyle {3\over 2}}$} at 21 2.5
\put {$\Omega$} at 23.5 2.0
\put {${\textstyle {1\over 2}}$} at 23 2.5
\put {${\textstyle {3\over 2}}$} at 24 2.5
%
%
%
\putrectangle corners at 0.6 5.20  and 1.4 5.55
\putrectangle corners at 0.6 6.40  and 1.4 6.80
\putrectangle corners at 1.6 5.15  and 2.4 5.30
\putrectangle corners at 1.6 6.50  and 2.4 7.50
\putrectangle corners at 2.6 6.70  and 3.4 6.85
\putrectangle corners at 4.6 6.15  and 5.4 6.75
\putrectangle corners at 5.6 6.70  and 6.4 7.70
\putrectangle corners at 7.6 7.30  and 8.4 8.00
\putrectangle corners at 8.6 6.65  and 9.4 6.85
\putrectangle corners at 8.6 9.00  and 9.4 9.50
\putrectangle corners at 9.6 7.70  and 10.4 7.80
\putrectangle corners at 11.6 4.03  and 12.4 4.11
\putrectangle corners at 11.6 6.60  and 12.4 6.80
\putrectangle corners at 11.6 7.20  and 12.4 8.50
\putrectangle corners at 12.6 5.185 and 13.4 5.195
\putrectangle corners at 12.6 6.85  and 13.4 6.95
\putrectangle corners at 13.6 8.10  and 14.4 8.30
\putrectangle corners at 16.6 8.18  and 17.4 8.28
%
%
%
%
\linethickness=2pt
\putrule from 0.6 5.29 to 1.4 5.29
\putrule from 0.6 6.31 to 1.4 6.31
\putrule from 1.6 5.20 to 2.4 5.20
\putrule from 1.6 7.52 to 2.4 7.52
\putrule from 2.6 6.74 to 3.4 6.74
\putrule from 4.6 8.84 to 5.4 8.84
\putrule from 5.6 7.57 to 6.4 7.57
\putrule from 7.6 7.15 to 8.4 7.15
\setdashpattern <3pt, 1pt>
\putrule from 7.6 8.40 to 8.4 8.40   
\putrule from 7.6 9.83 to 8.4 9.83   
\setsolid
\putrule from 8.6 7.04 to 9.4 7.04
\setdashpattern <3pt, 1pt>
\putrule from 8.6 8.76 to 9.4 8.76   
\setsolid
\putrule from 8.6 9.08 to 9.4 9.08
\putrule from 9.6 7.81 to 10.4 7.81
\putrule from 11.6 4.18 to 12.4 4.18
\putrule from 11.6 6.49 to 12.4 6.49
\putrule from 11.6 7.17 to 12.4 7.17
\putrule from 12.6 5.19 to 13.4 5.19
\putrule from 12.6 6.66 to 13.4 6.66
\setdashpattern <3pt, 1pt>
\putrule from 12.6 8.36 to 13.4 8.36   
\setsolid
\putrule from 13.6 8.34 to 14.4 8.34
\setdashpattern <3pt, 1pt>
\putrule from 15.6 8.06 to 16.4 8.06   
\putrule from 15.6 8.66 to 16.4 8.66   
\setsolid
\putrule from 16.6 8.19 to 17.4 8.19
\setdashpattern <3pt, 1pt>
\putrule from 16.6 9.55 to 17.4 9.55   
\putrule from 17.6 9.67 to 18.4 9.67   
\putrule from 19.6 10.76 to 20.4 10.76   
\putrule from 20.6 9.95 to 21.4 9.95   
\putrule from 22.6 11.72 to 23.4 11.72   
\putrule from 23.6 11.35 to 24.4 11.35   
\setsolid
\linethickness=0.7pt
\putrule from  18 4 to 18 5.75
\putrule from  18 5.75 to 25 5.75
\putrule from  25 5.75 to 25 4
\putrule from  25 4 to 18 4
\linethickness=2pt
\putrule from 18.5 5.25 to 19.3 5.25
\put {Fit Masses} [l] at 19.5 5.25
\setdashpattern <3pt, 1pt>
\putrule from 18.5 4.6 to 19.3 4.6
\setsolid
\put {Predicted Masses} [l] at 19.5 4.6
\endpicture$$

\vfill\eject

$$\beginpicture
\setcoordinatesystem units <0.225truein,0.5truein>
\setplotarea x from -1 to 25, y from 0 to 11
\putrule from 0 11 to 0 3
\putrule from 0 3 to 25 3
\putrule from -0.1 3 to 0 3
\putrule from -0.1 4 to 0 4
\putrule from -0.1 5 to 0 5
\putrule from -0.1 6 to 0 6
\putrule from -0.1 7 to 0 7
\putrule from -0.1 8 to 0 8
\putrule from -0.1 9 to 0 9
\putrule from -0.1 10 to 0 10
\putrule from -0.1 11 to 0 11
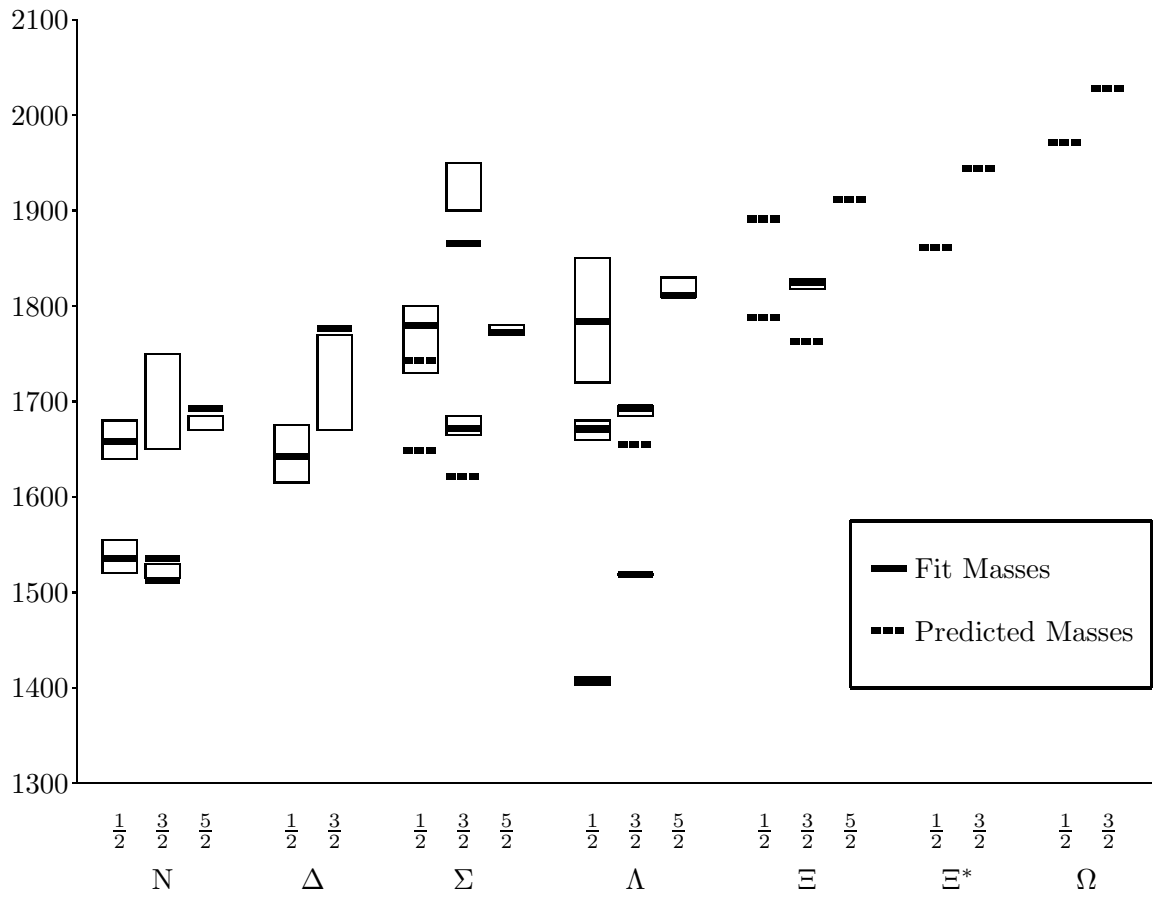
\put {Figure 3.  Masses of the $L=1$ Baryons in the Chiral Quark Model;
$\chi^2 = 24.7568$, $N_{fit} = 18$} [l] at -1 1.25
%
%
\put {1300} at -0.85 3 
\put {1400} at -0.85 4 
\put {1500} at -0.85 5 
\put {1600} at -0.85 6 
\put {1700} at -0.85 7 
\put {1800} at -0.85 8 
\put {1900} at -0.85 9 
\put {2000} at -0.85 10 
\put {2100} at -0.85 11 
%
%
%
\put {N} at 2 2.0
\put {${\textstyle {1\over 2}}$} at 1 2.5
\put {${\textstyle {3\over 2}}$} at 2 2.5
\put {${\textstyle {5\over 2}}$} at 3 2.5
\put {$\Delta$} at 5.5 2.0
\put {${\textstyle {1\over 2}}$} at 5 2.5
\put {${\textstyle {3\over 2}}$} at 6 2.5
\put {$\Sigma$} at 9 2.0
\put {${\textstyle {1\over 2}}$} at 8 2.5
\put {${\textstyle {3\over 2}}$} at 9 2.5
\put {${\textstyle {5\over 2}}$} at 10 2.5
\put {$\Lambda$} at 13 2.0
\put {${\textstyle {1\over 2}}$} at 12 2.5
\put {${\textstyle {3\over 2}}$} at 13 2.5
\put {${\textstyle {5\over 2}}$} at 14 2.5
\put {$\Xi$} at 17 2.0
\put {${\textstyle {1\over 2}}$} at 16 2.5
\put {${\textstyle {3\over 2}}$} at 17 2.5
\put {${\textstyle {5\over 2}}$} at 18 2.5
\put {$\Xi^*$} at 20.5 2.0
\put {${\textstyle {1\over 2}}$} at 20 2.5
\put {${\textstyle {3\over 2}}$} at 21 2.5
\put {$\Omega$} at 23.5 2.0
\put {${\textstyle {1\over 2}}$} at 23 2.5
\put {${\textstyle {3\over 2}}$} at 24 2.5
%
%
%
\putrectangle corners at 0.6 5.20  and 1.4 5.55
\putrectangle corners at 0.6 6.40  and 1.4 6.80
\putrectangle corners at 1.6 5.15  and 2.4 5.30
\putrectangle corners at 1.6 6.50  and 2.4 7.50
\putrectangle corners at 2.6 6.70  and 3.4 6.85
\putrectangle corners at 4.6 6.15  and 5.4 6.75
\putrectangle corners at 5.6 6.70  and 6.4 7.70
\putrectangle corners at 7.6 7.30  and 8.4 8.00
\putrectangle corners at 8.6 6.65  and 9.4 6.85
\putrectangle corners at 8.6 9.00  and 9.4 9.50
\putrectangle corners at 9.6 7.70  and 10.4 7.80
\putrectangle corners at 11.6 4.03  and 12.4 4.11
\putrectangle corners at 11.6 6.60  and 12.4 6.80
\putrectangle corners at 11.6 7.20  and 12.4 8.50
\putrectangle corners at 12.6 5.185 and 13.4 5.195
\putrectangle corners at 12.6 6.85  and 13.4 6.95
\putrectangle corners at 13.6 8.10  and 14.4 8.30
\putrectangle corners at 16.6 8.18  and 17.4 8.28
%
%
%
%
\linethickness=2pt
\putrule from 0.6 5.35 to 1.4 5.35
\putrule from 0.6 6.58 to 1.4 6.58
\putrule from 1.6 5.12 to 2.4 5.12
\putrule from 1.6 5.35 to 2.4 5.35
\putrule from 2.6 6.92 to 3.4 6.92
\putrule from 4.6 6.42 to 5.4 6.42
\putrule from 5.6 7.76 to 6.4 7.76
\setdashpattern <3pt, 1pt>
\putrule from 7.6 6.48 to 8.4 6.48   
\putrule from 7.6 7.43 to 8.4 7.43   
\setsolid
\putrule from 7.6 7.79 to 8.4 7.79
\setdashpattern <3pt, 1pt>
\putrule from 8.6 6.21 to 9.4 6.21   
\setsolid
\putrule from 8.6 6.72 to 9.4 6.72
\putrule from 8.6 8.65 to 9.4 8.65
\putrule from 9.6 7.72 to 10.4 7.72
\putrule from 11.6 4.08 to 12.4 4.08
\putrule from 11.6 6.71 to 12.4 6.71
\putrule from 11.6 7.84 to 12.4 7.84
\putrule from 12.6 5.19 to 13.4 5.19
\setdashpattern <3pt, 1pt>
\putrule from 12.6 6.55 to 13.4 6.55   
\setsolid
\putrule from 12.6 6.93 to 13.4 6.93
\putrule from 13.6 8.11 to 14.4 8.11
\setdashpattern <3pt, 1pt>
\putrule from 15.6 7.88 to 16.4 7.88   
\putrule from 15.6 8.91 to 16.4 8.91   
\putrule from 16.6 7.63 to 17.4 7.63   
\setsolid
\putrule from 16.6 8.24 to 17.4 8.24
\setdashpattern <3pt, 1pt>
\putrule from 17.6 9.11 to 18.4 9.11   
\putrule from 19.6 8.61 to 20.4 8.61   
\putrule from 20.6 9.44 to 21.4 9.44   
\putrule from 22.6 9.71 to 23.4 9.71   
\putrule from 23.6 10.28 to 24.4 10.28   
\setsolid%
\linethickness=0.7pt
\putrule from 18 4 to 18 5.75
\putrule from 18 5.75 to 25 5.75
\putrule from 25 5.75 to 25 4
\putrule from 25 4 to 18 4
\linethickness=2pt
\putrule from 18.5 5.25 to 19.3 5.25
\put {Fit Masses} [l] at 19.5 5.25
\setdashpattern <3pt, 1pt>
\putrule from 18.5 4.6 to 19.3 4.6
\setsolid
\put {Predicted Masses} [l] at 19.5 4.6
\endpicture$$

\vfill\eject

$$\beginpicture
\setcoordinatesystem units <0.750truein,0.250truein>
\setplotarea x from -1 to 7, y from -31 to 2.0
%
%

\put {Table 3.  Masses and Mixing Angles of the $L=1$ Baryons in the 
Quark Model.} [l] at -1 -32
%
%
\put {Baryon} [c] at -0.25 0.75 
\put {Mass} [c] at 1 1.25 
\put {(Exp)} [c] at 1 0.375 
\put {Mass} [c] at 2 1.25
\put {(Fit)} [c] at 2 0.375
\put {$^4$8} [c] at 3 0.75
\put {$^2$8} [c] at 4 0.75
\put {$^2$10} [c] at 5 0.75
\put {$^2$1} [c] at 6 0.75
\linethickness=1pt
\putrule from -1 1.75 to 6.5 1.75
\putrule from -1 -31 to 6.5 -31
\linethickness=1.5pt
\putrule from -1 -0.125 to 6.5 -0.125
%
\put {$N$} [c] at -0.625 -1
\put {$J=\half$} [c] at 0 -1
\put {$1535^{+20}_{-15}$} [c] at 1 -1
\put {1529} [c] at 2 -1
\put {-0.3537} [c] at 3 -1
\put {-0.9354} [c] at 4 -1
\put {*} [c] at 5 -1
\put {*} [c] at 6 -1
\put {$1650^{+30}_{-10}$} [c] at 1 -2
\put {1631} [c] at 2 -2
\put {-0.9354} [c] at 3 -2
\put {0.3537} [c] at 4 -2
\put {*} [c] at 5 -2
\put {*} [c] at 6 -2
%
\put {$N$} [c] at -0.625 -3
\put {$J=\thalf$} [c] at 0 -3
\put {$1520^{+10}_{-5}$} [c] at 1 -3
\put {1520} [c] at 2 -3
\put {-0.3283} [c] at 3 -3
\put {-0.9446} [c] at 4 -3
\put {*} [c] at 5 -3
\put {*} [c] at 6 -3
\put {$1700^{+50}_{-50}$} [c] at 1 -4
\put {1752} [c] at 2 -4
\put {-0.9446} [c] at 3 -4
\put {0.3283} [c] at 4 -4
\put {*} [c] at 5 -4
\put {*} [c] at 6 -4
%
\put {$N$} [c] at -0.625 -5
\put {$J=\fhalf$} [c] at 0 -5
\put {$1675^{+10}_{-5}$} [c] at 1 -5
\put {1674} [c] at 2 -5
\put {1.0000} [c] at 3 -5
\put {*} [c] at 4 -5
\put {*} [c] at 5 -5
\put {*} [c] at 6 -5
%
\put {$\Delta$} [c] at -0.625 -6
\put {$J=\half$} [c] at 0 -6
\put {$1620^{+55}_{-5}$} [c] at 1 -6
\put {1884} [c] at 2 -6
\put {*} [c] at 3 -6
\put {*} [c] at 4 -6
\put {1.0000} [c] at 5 -6
\put {*} [c] at 6 -6
%
\put {$\Delta$} [c] at -0.625 -7
\put {$J=\thalf$} [c] at 0 -7
\put {$1700^{+70}_{-30}$} [c] at 1 -7
\put {1757} [c] at 2 -7
\put {*} [c] at 3 -7
\put {*} [c] at 4 -7
\put {1.0000} [c] at 5 -7
\put {*} [c] at 6 -7
%
\put {$\Sigma$} [c] at -0.625 -8
\put {$J=\half$} [c] at 0 -8
\put {$1750^{+50}_{-20}$} [c] at 1 -8
\put {1715} [c] at 2 -8
\put {-0.1874} [c] at 3 -8
\put {-0.9773} [c] at 4 -8
\put {0.0988} [c] at 5 -8
\put {*} [c] at 6 -8
\put {**} [c] at 1 -9
\put {1840} [c] at 2 -9
\put {0.9819} [c] at 3 -9
\put {-0.1892} [c] at 4 -9
\put {-0.0094} [c] at 5 -9
\put {*} [c] at 6 -9
\put {**} [c] at 1 -10
\put {1983} [c] at 2 -10
\put {0.0279} [c] at 3 -10
\put {0.0952} [c] at 4 -10
\put {0.9951} [c] at 5 -10
\put {*} [c] at 6 -10
%
\put {$\Sigma$} [c] at -0.625 -11
\put {$J=\thalf$} [c] at 0 -11
\put {$1670^{+15}_{-5}$} [c] at 1 -11
\put {1704} [c] at 2 -11
\put {0.3113} [c] at 3 -11
\put {0.9480} [c] at 4 -11
\put {-0.0658} [c] at 5 -11
\put {*} [c] at 6 -11
\put {**} [c] at 1 -12
\put {1876} [c] at 2 -12
\put {0.0803} [c] at 3 -12
\put {0.0428} [c] at 4 -12
\put {0.9959} [c] at 5 -12
\put {*} [c] at 6 -12
\put {$1940^{+10}_{-40}$} [c] at 1 -13
\put {1908} [c] at 2 -13
\put {0.9469} [c] at 3 -13
\put {-0.3153} [c] at 4 -13
\put {-0.0628} [c] at 5 -13
\put {*} [c] at 6 -13
%
\put {$\Sigma$} [c] at -0.625 -14
\put {$J=\fhalf$} [c] at 0 -14
\put {$1775^{+5}_{-5}$} [c] at 1 -14
\put {1781} [c] at 2 -14
\put {1.0000} [c] at 3 -14
\put {*} [c] at 4 -14
\put {*} [c] at 5 -14
\put {*} [c] at 6 -14
%
\put {$\Lambda$} [c] at -0.625 -15
\put {$J=\half$} [c] at 0 -15
\put {$1407^{+4}_{-4}$} [c] at 1 -15
\put {1418} [c] at 2 -15
\put {0.01780} [c] at 3 -15
\put {0.0082} [c] at 4 -15
\put {*} [c] at 5 -15
\put {-0.9998} [c] at 6 -15
\put {$1670^{+10}_{-10}$} [c] at 1 -16
\put {1649} [c] at 2 -16
\put {0.6359} [c] at 3 -16
\put {0.7716} [c] at 4 -16
\put {*} [c] at 5 -16
\put {0.01767} [c] at 6 -16
\put {$1800^{+50}_{-80}$} [c] at 1 -17
\put {1717} [c] at 2 -17
\put {0.7716} [c] at 3 -17
\put {-0.6361} [c] at 4 -17
\put {*} [c] at 5 -17
\put {0.0085} [c] at 6 -17
%
\put {$\Lambda$} [c] at -0.625 -18
\put {$J=\thalf$} [c] at 0 -18
\put {$1519.5^{+1}_{-1}$} [c] at 1 -18
\put {1519} [c] at 2 -18
\put {0.0503} [c] at 3 -18
\put {0.2395} [c] at 4 -18
\put {*} [c] at 5 -18
\put {-0.9696} [c] at 6 -18
\put {$1690^{+5}_{-5}$} [c] at 1 -19
\put {1666} [c] at 2 -19
\put {0.3513} [c] at 3 -19
\put {0.9045} [c] at 4 -19
\put {*} [c] at 5 -19
\put {0.2416} [c] at 6 -19
\put {**} [c] at 1 -20
\put {1836} [c] at 2 -20
\put {0.9349} [c] at 3 -20
\put {-0.3528} [c] at 4 -20
\put {*} [c] at 5 -20
\put {-0.0386} [c] at 6 -20
%
\put {$\Lambda$} [c] at -0.625 -21
\put {$J=\fhalf$} [c] at 0 -21
\put {$1830^{+0}_{-20}$} [c] at 1 -21
\put {1834} [c] at 2 -21
\put {1.0000} [c] at 3 -21
\put {*} [c] at 4 -21
\put {*} [c] at 5 -21
\put {*} [c] at 6 -21
%
\put {$\Xi$} [c] at -0.625 -22
\put {$J=\half$} [c] at 0 -22
\put {**} [c] at 1 -22
\put {1806} [c] at 2 -22
\put {0.7132} [c] at 3 -22
\put {0.7010} [c] at 4 -22
\put {*} [c] at 5 -22
\put {*} [c] at 6 -22
\put {**} [c] at 1 -23
\put {1860} [c] at 2 -23
\put {-0.7010} [c] at 3 -23
\put {0.7132} [c] at 4 -23
\put {*} [c] at 5 -23
\put {*} [c] at 6 -23
%
\put {$\Xi$} [c] at -0.625 -24
\put {$J=\thalf$} [c] at 0 -24
\put {$1823^{+5}_{-5}$} [c] at 1 -24
\put {1819} [c] at 2 -24
\put {-0.3514} [c] at 3 -24
\put {-0.9362} [c] at 4 -24
\put {*} [c] at 5 -24
\put {*} [c] at 6 -24
\put {**} [c] at 1 -25
\put {1955} [c] at 2 -25
\put {-0.9362} [c] at 3 -25
\put {0.3514} [c] at 4 -25
\put {*} [c] at 5 -25
\put {*} [c] at 6 -25
%
\put {$\Xi$} [c] at -0.625 -26
\put {$J=\fhalf$} [c] at 0 -26
\put {**} [c] at 1 -26
\put {1967} [c] at 2 -26
\put {1.0000} [c] at 3 -26
\put {*} [c] at 4 -26
\put {*} [c] at 5 -26
\put {*} [c] at 6 -26
%
\put {$\Xi^*$} [c] at -0.625 -27
\put {$J=\half$} [c] at 0 -27
\put {**} [c] at 1 -27
\put {2076} [c] at 2 -27
\put {*} [c] at 3 -27
\put {*} [c] at 4 -27
\put {1.0000} [c] at 5 -27
\put {*} [c] at 6 -27
%
\put {$\Xi^*$} [c] at -0.625 -28
\put {$J=\thalf$} [c] at 0 -28
\put {**} [c] at 1 -28
\put {1995} [c] at 2 -28
\put {*} [c] at 3 -28
\put {*} [c] at 4 -28
\put {1.0000} [c] at 5 -28
\put {*} [c] at 6 -28
%
\put {$\Omega$} [c] at -0.625 -29
\put {$J=\half$} [c] at 0 -29
\put {**} [c] at 1 -29
\put {2172} [c] at 2 -29
\put {*} [c] at 3 -29
\put {*} [c] at 4 -29
\put {1.0000} [c] at 5 -29
\put {*} [c] at 6 -29
%
\put {$\Omega$} [c] at -0.625 -30
\put {$J=\thalf$} [c] at 0 -30
\put {**} [c] at 1 -30
\put {2113} [c] at 2 -30
\put {*} [c] at 3 -30
\put {*} [c] at 4 -30
\put {1.0000} [c] at 5 -30
\put {*} [c] at 6 -30
\endpicture$$

\vfill\eject

$$\beginpicture
\setcoordinatesystem units <0.750truein,0.250truein>
\setplotarea x from -1 to 7, y from -31 to 2.0
%
%
\put {Table 4.  Masses and Mixing Angles of the $L=1$ Baryons 
in the Chiral Quark Model.} [l] at -1 -32
%
%
\put {Baryon} [c] at -0.25 0.75 
\put {Mass} [c] at 1 1.25 
\put {(Exp)} [c] at 1 0.375 
\put {Mass} [c] at 2 1.25
\put {(Fit)} [c] at 2 0.375
\put {$^4$8} [c] at 3 0.75
\put {$^2$8} [c] at 4 0.75
\put {$^2$10} [c] at 5 0.75
\put {$^2$1} [c] at 6 0.75
\linethickness=1pt
\putrule from -1 1.75 to 6.5 1.75
\putrule from -1 -31 to 6.5 -31
\linethickness=1.5pt
\putrule from -1 -0.125 to 6.5 -0.125
%
\put {$N$} [c] at -0.625 -1
\put {$J=\half$} [c] at 0 -1
\put {$1535^{+20}_{-15}$} [c] at 1 -1
\put {1536} [c] at 2 -1
\put {-0.4849} [c] at 3 -1
\put {-0.8746} [c] at 4 -1
\put {*} [c] at 5 -1
\put {*} [c] at 6 -1
\put {$1650^{+30}_{-10}$} [c] at 1 -2
\put {1658} [c] at 2 -2
\put {-0.8746} [c] at 3 -2
\put {0.4849} [c] at 4 -2
\put {*} [c] at 5 -2
\put {*} [c] at 6 -2
%
\put {$N$} [c] at -0.625 -3
\put {$J=\thalf$} [c] at 0 -3
\put {$1520^{+10}_{-5}$} [c] at 1 -3
\put {1512} [c] at 2 -3
\put {0.4915} [c] at 3 -3
\put {-0.8709} [c] at 4 -3
\put {*} [c] at 5 -3
\put {*} [c] at 6 -3
\put {$1700^{+50}_{-50}$} [c] at 1 -4
\put {1535} [c] at 2 -4
\put {-0.8709} [c] at 3 -4
\put {-0.4915} [c] at 4 -4
\put {*} [c] at 5 -4
\put {*} [c] at 6 -4
%
\put {$N$} [c] at -0.625 -5
\put {$J=\fhalf$} [c] at 0 -5
\put {$1675^{+10}_{-5}$} [c] at 1 -5
\put {1692} [c] at 2 -5
\put {1.0000} [c] at 3 -5
\put {*} [c] at 4 -5
\put {*} [c] at 5 -5
\put {*} [c] at 6 -5
%
\put {$\Delta$} [c] at -0.625 -6
\put {$J=\half$} [c] at 0 -6
\put {$1620^{+55}_{-5}$} [c] at 1 -6
\put {1643} [c] at 2 -6
\put {*} [c] at 3 -6
\put {*} [c] at 4 -6
\put {1.0000} [c] at 5 -6
\put {*} [c] at 6 -6
%
\put {$\Delta$} [c] at -0.625 -7
\put {$J=\thalf$} [c] at 0 -7
\put {$1700^{+70}_{-30}$} [c] at 1 -7
\put {1776} [c] at 2 -7
\put {*} [c] at 3 -7
\put {*} [c] at 4 -7
\put {1.0000} [c] at 5 -7
\put {*} [c] at 6 -7
%
\put {$\Sigma$} [c] at -0.625 -8
\put {$J=\half$} [c] at 0 -8
\put {$(1620)$} [c] at 1 -8
\put {1648} [c] at 2 -8
\put {-0.5882} [c] at 3 -8
\put {-0.7248} [c] at 4 -8
\put {-0.3588} [c] at 5 -8
\put {*} [c] at 6 -8
\put {**} [c] at 1 -9
\put {1743} [c] at 2 -9
\put {0.7688} [c] at 3 -9
\put {-0.3636} [c] at 4 -9
\put {-0.5260} [c] at 5 -9
\put {*} [c] at 6 -9
\put {$1750^{+50}_{-20}$} [c] at 1 -10
\put {1779} [c] at 2 -10
\put {-0.2508} [c] at 3 -10
\put {0.5853} [c] at 4 -10
\put {-0.7711} [c] at 5 -10
\put {*} [c] at 6 -10
%
\put {$\Sigma$} [c] at -0.625 -11
\put {$J=\thalf$} [c] at 0 -11
\put {$(1580)$} [c] at 1 -11
\put {1621} [c] at 2 -11
\put {0.9320} [c] at 3 -11
\put {-0.3575} [c] at 4 -11
\put {0.0604} [c] at 5 -11
\put {*} [c] at 6 -11
\put {$1670^{+15}_{-5}$} [c] at 1 -12
\put {1672} [c] at 2 -12
\put {-0.3625} [c] at 3 -12
\put {-0.9192} [c] at 4 -12
\put {0.1538} [c] at 5 -12
\put {*} [c] at 6 -12
\put {$1940^{+10}_{-40}$} [c] at 1 -13
\put {1865} [c] at 2 -13
\put {-0.0005} [c] at 3 -13
\put {0.1652} [c] at 4 -13
\put {0.9863} [c] at 5 -13
\put {*} [c] at 6 -13
%
\put {$\Sigma$} [c] at -0.625 -14
\put {$J=\fhalf$} [c] at 0 -14
\put {$1775^{+5}_{-5}$} [c] at 1 -14
\put {1772} [c] at 2 -14
\put {1.0000} [c] at 3 -14
\put {*} [c] at 4 -14
\put {*} [c] at 5 -14
\put {*} [c] at 6 -14
%
\put {$\Lambda$} [c] at -0.625 -15
\put {$J=\half$} [c] at 0 -15
\put {$1407^{+4}_{-4}$} [c] at 1 -15
\put {1408} [c] at 2 -15
\put {-0.0395} [c] at 3 -15
\put {-0.2071} [c] at 4 -15
\put {*} [c] at 5 -15
\put {-0.9775} [c] at 6 -15
\put {$1670^{+10}_{-10}$} [c] at 1 -16
\put {1671} [c] at 2 -16
\put {0.3384} [c] at 3 -16
\put {0.9177} [c] at 4 -16
\put {*} [c] at 5 -16
\put {-0.2081} [c] at 6 -16
\put {$1800^{+50}_{-80}$} [c] at 1 -17
\put {1784} [c] at 2 -17
\put {0.9402} [c] at 3 -17
\put {-0.3391} [c] at 4 -17
\put {*} [c] at 5 -17
\put {0.0339} [c] at 6 -17
%
\put {$\Lambda$} [c] at -0.625 -18
\put {$J=\thalf$} [c] at 0 -18
\put {$1519.5^{+1}_{-1}$} [c] at 1 -18
\put {1519} [c] at 2 -18
\put {0.0243} [c] at 3 -18
\put {0.4288} [c] at 4 -18
\put {*} [c] at 5 -18
\put {-0.9031} [c] at 6 -18
\put {**} [c] at 1 -19
\put {1655} [c] at 2 -19
\put {-0.0240} [c] at 3 -19
\put {0.9033} [c] at 4 -19
\put {*} [c] at 5 -19
\put {0.4283} [c] at 6 -19
\put {$1690^{+5}_{-5}$} [c] at 1 -20
\put {1693} [c] at 2 -20
\put {0.9994} [c] at 3 -20
\put {0.0113} [c] at 4 -20
\put {*} [c] at 5 -20
\put {0.0322} [c] at 6 -20
%
\put {$\Lambda$} [c] at -0.625 -21
\put {$J=\fhalf$} [c] at 0 -21
\put {$1830^{+0}_{-20}$} [c] at 1 -21
\put {1811} [c] at 2 -21
\put {1.0000} [c] at 3 -21
\put {*} [c] at 4 -21
\put {*} [c] at 5 -21
\put {*} [c] at 6 -21
%
\put {$\Xi$} [c] at -0.625 -22
\put {$J=\half$} [c] at 0 -22
\put {**} [c] at 1 -22
\put {1788} [c] at 2 -22
\put {-0.2884} [c] at 3 -22
\put {-0.9575} [c] at 4 -22
\put {*} [c] at 5 -22
\put {*} [c] at 6 -22
\put {**} [c] at 1 -23
\put {1891} [c] at 2 -23
\put {-0.9575} [c] at 3 -23
\put {0.2884} [c] at 4 -23
\put {*} [c] at 5 -23
\put {*} [c] at 6 -23
%
\put {$\Xi$} [c] at -0.625 -24
\put {$J=\thalf$} [c] at 0 -24
\put {**} [c] at 1 -24
\put {1763} [c] at 2 -24
\put {-0.0477} [c] at 3 -24
\put {-0.9989} [c] at 4 -24
\put {*} [c] at 5 -24
\put {*} [c] at 6 -24
\put {$1823^{+5}_{-5}$} [c] at 1 -25
\put {1824} [c] at 2 -25
\put {-0.9989} [c] at 3 -25
\put {0.0477} [c] at 4 -25
\put {*} [c] at 5 -25
\put {*} [c] at 6 -25
%
\put {$\Xi$} [c] at -0.625 -26
\put {$J=\fhalf$} [c] at 0 -26
\put {**} [c] at 1 -26
\put {1911} [c] at 2 -26
\put {1.0000} [c] at 3 -26
\put {*} [c] at 4 -26
\put {*} [c] at 5 -26
\put {*} [c] at 6 -26
%
\put {$\Xi^*$} [c] at -0.625 -27
\put {$J=\half$} [c] at 0 -27
\put {**} [c] at 1 -27
\put {1861} [c] at 2 -27
\put {*} [c] at 3 -27
\put {*} [c] at 4 -27
\put {1.0000} [c] at 5 -27
\put {*} [c] at 6 -27
%
\put {$\Xi^*$} [c] at -0.625 -28
\put {$J=\thalf$} [c] at 0 -28
\put {**} [c] at 1 -28
\put {1944} [c] at 2 -28
\put {*} [c] at 3 -28
\put {*} [c] at 4 -28
\put {1.0000} [c] at 5 -28
\put {*} [c] at 6 -28
%
\put {$\Omega$} [c] at -0.625 -29
\put {$J=\half$} [c] at 0 -29
\put {**} [c] at 1 -29
\put {1971} [c] at 2 -29
\put {*} [c] at 3 -29
\put {*} [c] at 4 -29
\put {1.0000} [c] at 5 -29
\put {*} [c] at 6 -29
%
\put {$\Omega$} [c] at -0.625 -30
\put {$J=\thalf$} [c] at 0 -30
\put {**} [c] at 1 -30
\put {2028} [c] at 2 -30
\put {*} [c] at 3 -30
\put {*} [c] at 4 -30
\put {1.0000} [c] at 5 -30
\put {*} [c] at 6 -30
\endpicture$$

\vfill\eject

$$\beginpicture
\setcoordinatesystem units <0.5truein,0.25truein>
\setplotarea x from -1 to 5, y from -15 to 2.5
\linethickness=1.0pt
\putrule from -1 2.25 to 5 2.25
\linethickness=1.5pt
\putrule from -1 0 to 5 0
\linethickness=1.0pt
\putrule from -1 -11 to 5 -11
\putrule from 1 -11 to 1 0
\put {Perturbation} [c] at 0 1.5 
\put {(MeV)$^*$} [c] at 0 0.625 
\put {Quark} [c] at 2 1.5
\put {Model} [c] at 2 0.625
\put {Chiral} [c] at 4 1.5
\put {Quark Model} [c] at 4 0.625
\put {${\cal D}_+$} [c] at 0 -1
\put {${\cal D}_-$} [c] at 0 -2
\put {${\cal SO}_1$} [c] at 0 -3
\put {${\cal SO}_2$} [c] at 0 -4
\put {${\cal SO}_3$} [c] at 0 -5
\put {${\cal Q}_1$} [c] at 0 -6
\put {${\cal Q}_2$} [c] at 0 -7
\put {$\delta m$} [c] at 0 -8
\put {$M_0$} [c] at 0 -9
\put {$\chi^2$} [c] at 0 -10
\put {123.9} [c] at 2 -1
\put {-21.14} [c] at 2 -2
\put {-14.32} [c] at 2 -3
\put {-77.90} [c] at 2 -4
\put {68.34} [c] at 2 -5
\put {-18.20} [c] at 2 -6
\put {-10.12} [c] at 2 -7
\put {0.221} [c] at 2 -8
\put {1585} [c] at 2 -9
\put {79.24} [c] at 2 -10
\put {-79.73} [c] at 4 -1
\put {-57.05} [c] at 4 -2
\put {0.744} [c] at 4 -3
\put {16.24} [c] at 4 -4
\put {20.38} [c] at 4 -5
\put {2.699} [c] at 4 -6
\put {-0.982} [c] at 4 -7
\put {0.265} [c] at 4 -8
\put {1445} [c] at 4 -9
\put {33.23} [c] at 4 -10
\put {Table 5.  Best-fit values of the $S_3$ constants for} [l] at -1 -12 
\put {the quark model and chiral quark model with} [l] at -1 -12.8
\put {$SU(3)$ symmetric spin operators.  $^*$All of the} [l] at -1 -13.8
\put {parameters are in units of MeV except $\delta m$} [l] at -1 -14.6
\put {which is dimensionless.} [l] at -1 -15.4
\endpicture$$

\vfill\eject

$$\beginpicture
\setcoordinatesystem units <0.225truein,0.5truein>
\setplotarea x from -1 to 25, y from 0 to 12
\putrule from 0 12 to 0 3
\putrule from 0 3 to 25 3
\putrule from -0.1 3 to 0 3
\putrule from -0.1 4 to 0 4
\putrule from -0.1 5 to 0 5
\putrule from -0.1 6 to 0 6
\putrule from -0.1 7 to 0 7
\putrule from -0.1 8 to 0 8
\putrule from -0.1 9 to 0 9
\putrule from -0.1 10 to 0 10
\putrule from -0.1 11 to 0 11
\putrule from -0.1 12 to 0 12
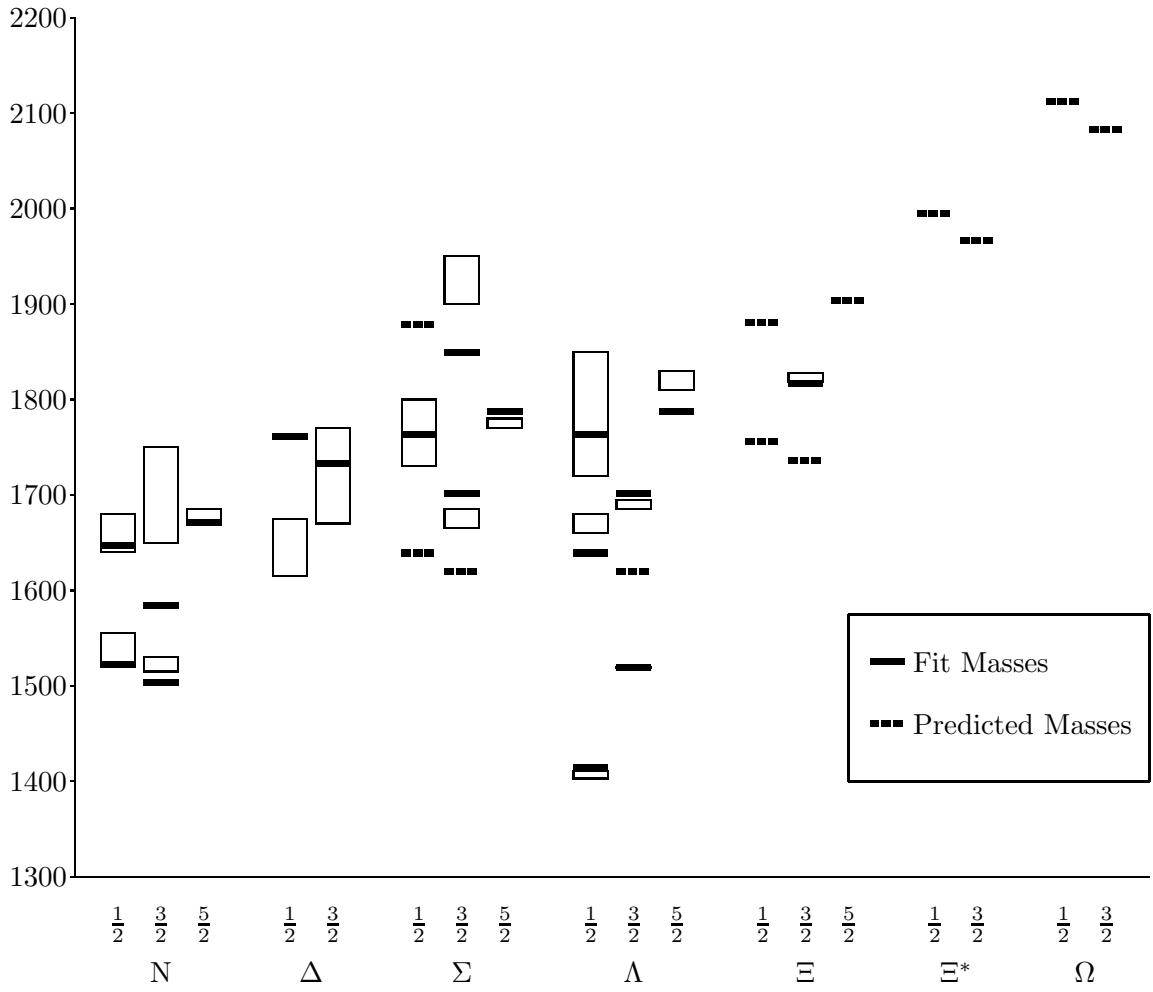
\put {Figure 4.  Masses of the $L=1$ Baryons in the Quark Model with
$SU(3)$ Symmetric } [l] at -1 1.25
\put {Interactions; $\chi^2 = 79.2456$, $N_{fit} = 18$} [l] at -1 0.75
%
%
\put {1300} at -0.85 3 
\put {1400} at -0.85 4 
\put {1500} at -0.85 5 
\put {1600} at -0.85 6 
\put {1700} at -0.85 7 
\put {1800} at -0.85 8 
\put {1900} at -0.85 9 
\put {2000} at -0.85 10 
\put {2100} at -0.85 11 
\put {2200} at -0.85 12 
%
%
%
\put {N} at 2 2.0
\put {${\textstyle {1\over 2}}$} at 1 2.5
\put {${\textstyle {3\over 2}}$} at 2 2.5
\put {${\textstyle {5\over 2}}$} at 3 2.5
\put {$\Delta$} at 5.5 2.0
\put {${\textstyle {1\over 2}}$} at 5 2.5
\put {${\textstyle {3\over 2}}$} at 6 2.5
\put {$\Sigma$} at 9 2.0
\put {${\textstyle {1\over 2}}$} at 8 2.5
\put {${\textstyle {3\over 2}}$} at 9 2.5
\put {${\textstyle {5\over 2}}$} at 10 2.5
\put {$\Lambda$} at 13 2.0
\put {${\textstyle {1\over 2}}$} at 12 2.5
\put {${\textstyle {3\over 2}}$} at 13 2.5
\put {${\textstyle {5\over 2}}$} at 14 2.5
\put {$\Xi$} at 17 2.0
\put {${\textstyle {1\over 2}}$} at 16 2.5
\put {${\textstyle {3\over 2}}$} at 17 2.5
\put {${\textstyle {5\over 2}}$} at 18 2.5
\put {$\Xi^*$} at 20.5 2.0
\put {${\textstyle {1\over 2}}$} at 20 2.5
\put {${\textstyle {3\over 2}}$} at 21 2.5
\put {$\Omega$} at 23.5 2.0
\put {${\textstyle {1\over 2}}$} at 23 2.5
\put {${\textstyle {3\over 2}}$} at 24 2.5
%
%
%
\putrectangle corners at 0.6 5.20  and 1.4 5.55
\putrectangle corners at 0.6 6.40  and 1.4 6.80
\putrectangle corners at 1.6 5.15  and 2.4 5.30
\putrectangle corners at 1.6 6.50  and 2.4 7.50
\putrectangle corners at 2.6 6.70  and 3.4 6.85
\putrectangle corners at 4.6 6.15  and 5.4 6.75
\putrectangle corners at 5.6 6.70  and 6.4 7.70
\putrectangle corners at 7.6 7.30  and 8.4 8.00
\putrectangle corners at 8.6 6.65  and 9.4 6.85
\putrectangle corners at 8.6 9.00  and 9.4 9.50
\putrectangle corners at 9.6 7.70  and 10.4 7.80
\putrectangle corners at 11.6 4.03  and 12.4 4.11
\putrectangle corners at 11.6 6.60  and 12.4 6.80
\putrectangle corners at 11.6 7.20  and 12.4 8.50
\putrectangle corners at 12.6 5.185 and 13.4 5.195
\putrectangle corners at 12.6 6.85  and 13.4 6.95
\putrectangle corners at 13.6 8.10  and 14.4 8.30
\putrectangle corners at 16.6 8.18  and 17.4 8.28
%
%
%
%
\linethickness=2pt
\putrule from 0.6 5.22 to 1.4 5.22
\putrule from 0.6 6.47 to 1.4 6.47
\putrule from 1.6 5.03 to 2.4 5.03
\putrule from 1.6 5.84 to 2.4 5.84
\putrule from 2.6 6.71 to 3.4 6.71
\putrule from 4.6 7.61 to 5.4 7.61
\putrule from 5.6 7.33 to 6.4 7.33
\setdashpattern <3pt, 1pt>
\putrule from 7.6 6.39 to 8.4 6.39   
\putrule from 7.6 8.78 to 8.4 8.78   
\setsolid
\putrule from 7.6 7.63 to 8.4 7.63
\setdashpattern <3pt, 1pt>
\putrule from 8.6 6.20 to 9.4 6.20   
\setsolid
\putrule from 8.6 7.01 to 9.4 7.01
\putrule from 8.6 8.49 to 9.4 8.49
\putrule from 9.6 7.87 to 10.4 7.87
\putrule from 11.6 4.14 to 12.4 4.14
\putrule from 11.6 6.39 to 12.4 6.39
\putrule from 11.6 7.63 to 12.4 7.63
\putrule from 12.6 5.19 to 13.4 5.19
\setdashpattern <3pt, 1pt>
\putrule from 12.6 6.20 to 13.4 6.20   
\setsolid
\putrule from 12.6 7.01 to 13.4 7.01
\putrule from 13.6 7.87 to 14.4 7.87
\setdashpattern <3pt, 1pt>
\putrule from 15.6 7.56 to 16.4 7.56   
\putrule from 15.6 8.80 to 16.4 8.80   
\putrule from 16.6 7.36 to 17.4 7.36   
\setsolid
\putrule from 16.6 8.17 to 17.4 8.17
\setdashpattern <3pt, 1pt>
\putrule from 17.6 9.04 to 18.4 9.04   
\putrule from 19.6 9.95 to 20.4 9.95   
\putrule from 20.6 9.66 to 21.4 9.66   
\putrule from 22.6 11.12 to 23.4 11.12   
\putrule from 23.6 10.83 to 24.4 10.83   
\setsolid
\linethickness=0.7pt
\putrule from 18 4 to 18 5.75
\putrule from 18 5.75 to 25 5.75
\putrule from 25 5.75 to 25 4
\putrule from 25 4 to 18 4
\linethickness=2pt
\putrule from 18.5 5.25 to 19.3 5.25
\put {Fit Masses} [l] at 19.5 5.25
\setdashpattern <3pt, 1pt>
\putrule from 18.5 4.6 to 19.3 4.6
\setsolid
\put {Predicted Masses} [l] at 19.5 4.6
\endpicture$$

\vfill\eject

$$\beginpicture
\setcoordinatesystem units <0.225truein,0.5truein>
\setplotarea x from -1 to 25, y from 0 to 12
\putrule from 0 12 to 0 3
\putrule from 0 3 to 25 3
\putrule from -0.1 3 to 0 3
\putrule from -0.1 4 to 0 4
\putrule from -0.1 5 to 0 5
\putrule from -0.1 6 to 0 6
\putrule from -0.1 7 to 0 7
\putrule from -0.1 8 to 0 8
\putrule from -0.1 9 to 0 9
\putrule from -0.1 10 to 0 10
\putrule from -0.1 11 to 0 11
\putrule from -0.1 12 to 0 12
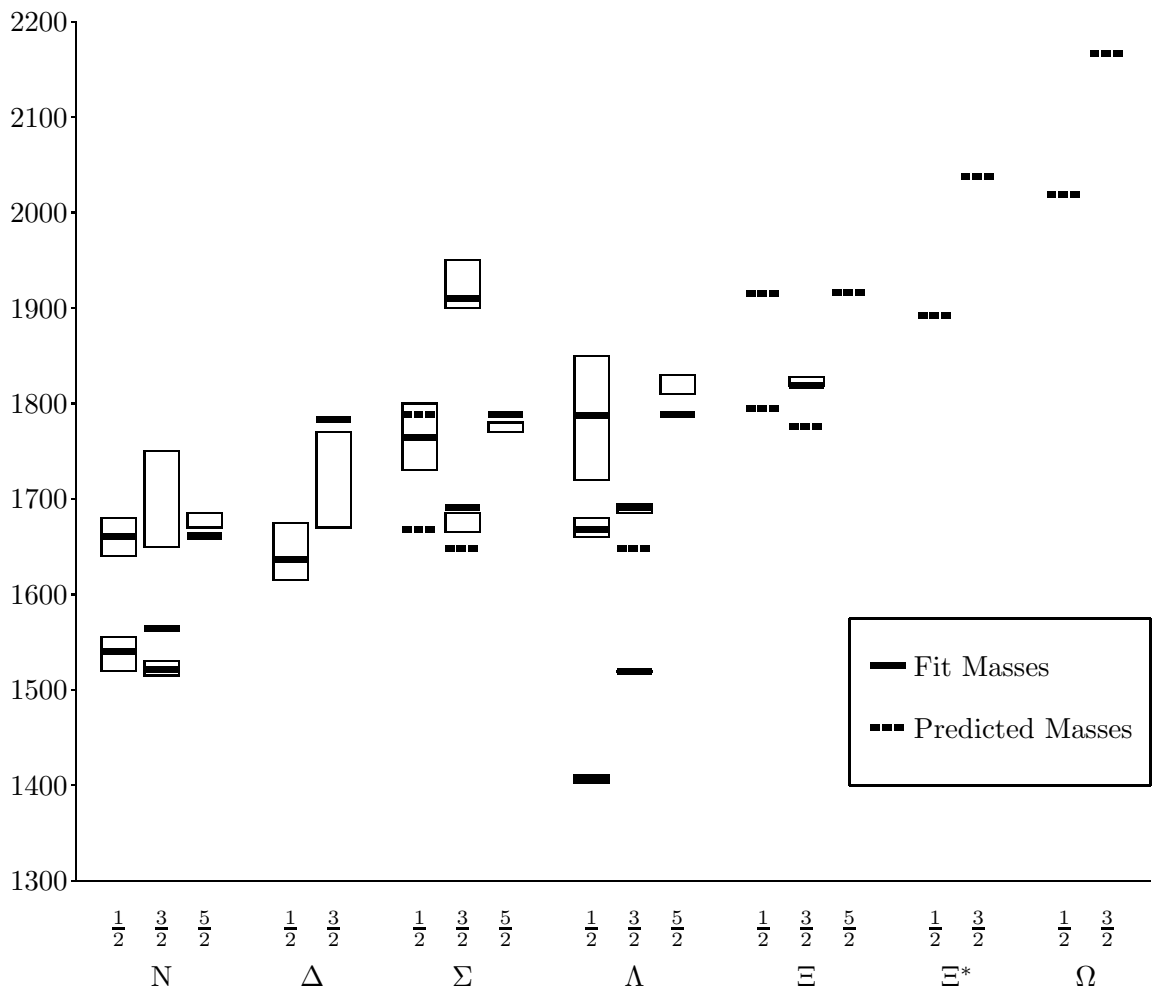
\put {Figure 5.  Masses of the $L=1$ Baryons in the Chiral Quark Model
with $SU(3)$ Symmetric} [l] at -1 1.25
\put {Interactions; $\chi^2 = 33.2254$, $N_{fit} = 18$} [l] at -1 0.75
%
%
\put {1300} at -0.85 3 
\put {1400} at -0.85 4 
\put {1500} at -0.85 5 
\put {1600} at -0.85 6 
\put {1700} at -0.85 7 
\put {1800} at -0.85 8 
\put {1900} at -0.85 9 
\put {2000} at -0.85 10 
\put {2100} at -0.85 11 
\put {2200} at -0.85 12 
%
%
%
\put {N} at 2 2.0
\put {${\textstyle {1\over 2}}$} at 1 2.5
\put {${\textstyle {3\over 2}}$} at 2 2.5
\put {${\textstyle {5\over 2}}$} at 3 2.5
\put {$\Delta$} at 5.5 2.0
\put {${\textstyle {1\over 2}}$} at 5 2.5
\put {${\textstyle {3\over 2}}$} at 6 2.5
\put {$\Sigma$} at 9 2.0
\put {${\textstyle {1\over 2}}$} at 8 2.5
\put {${\textstyle {3\over 2}}$} at 9 2.5
\put {${\textstyle {5\over 2}}$} at 10 2.5
\put {$\Lambda$} at 13 2.0
\put {${\textstyle {1\over 2}}$} at 12 2.5
\put {${\textstyle {3\over 2}}$} at 13 2.5
\put {${\textstyle {5\over 2}}$} at 14 2.5
\put {$\Xi$} at 17 2.0
\put {${\textstyle {1\over 2}}$} at 16 2.5
\put {${\textstyle {3\over 2}}$} at 17 2.5
\put {${\textstyle {5\over 2}}$} at 18 2.5
\put {$\Xi^*$} at 20.5 2.0
\put {${\textstyle {1\over 2}}$} at 20 2.5
\put {${\textstyle {3\over 2}}$} at 21 2.5
\put {$\Omega$} at 23.5 2.0
\put {${\textstyle {1\over 2}}$} at 23 2.5
\put {${\textstyle {3\over 2}}$} at 24 2.5
%
%
%
\putrectangle corners at 0.6 5.20  and 1.4 5.55
\putrectangle corners at 0.6 6.40  and 1.4 6.80
\putrectangle corners at 1.6 5.15  and 2.4 5.30
\putrectangle corners at 1.6 6.50  and 2.4 7.50
\putrectangle corners at 2.6 6.70  and 3.4 6.85
\putrectangle corners at 4.6 6.15  and 5.4 6.75
\putrectangle corners at 5.6 6.70  and 6.4 7.70
\putrectangle corners at 7.6 7.30  and 8.4 8.00
\putrectangle corners at 8.6 6.65  and 9.4 6.85
\putrectangle corners at 8.6 9.00  and 9.4 9.50
\putrectangle corners at 9.6 7.70  and 10.4 7.80
\putrectangle corners at 11.6 4.03  and 12.4 4.11
\putrectangle corners at 11.6 6.60  and 12.4 6.80
\putrectangle corners at 11.6 7.20  and 12.4 8.50
\putrectangle corners at 12.6 5.185 and 13.4 5.195
\putrectangle corners at 12.6 6.85  and 13.4 6.95
\putrectangle corners at 13.6 8.10  and 14.4 8.30
\putrectangle corners at 16.6 8.18  and 17.4 8.28
%
%
%
%
\linethickness=2pt
\putrule from 0.6 5.40 to 1.4 5.40
\putrule from 0.6 6.60 to 1.4 6.60
\putrule from 1.6 5.21 to 2.4 5.21
\putrule from 1.6 5.64 to 2.4 5.64
\putrule from 2.6 6.61 to 3.4 6.61
\putrule from 4.6 6.36 to 5.4 6.36
\putrule from 5.6 7.83 to 6.4 7.83
\setdashpattern <3pt, 1pt>
\putrule from 7.6 6.68 to 8.4 6.68   
\putrule from 7.6 7.88 to 8.4 7.88   
\setsolid
\putrule from 7.6 7.64 to 8.4 7.64
\setdashpattern <3pt, 1pt>
\putrule from 8.6 6.48 to 9.4 6.48   
\setsolid
\putrule from 8.6 6.91 to 9.4 6.91
\putrule from 8.6 9.10 to 9.4 9.10
\putrule from 9.6 7.88 to 10.4 7.88
\putrule from 11.6 4.07 to 12.4 4.07
\putrule from 11.6 6.68 to 12.4 6.68
\putrule from 11.6 7.87 to 12.4 7.87
\putrule from 12.6 5.19 to 13.4 5.19
\setdashpattern <3pt, 1pt>
\putrule from 12.6 6.48 to 13.4 6.48   
\setsolid
\putrule from 12.6 6.91 to 13.4 6.91
\putrule from 13.6 7.88 to 14.4 7.88
\setdashpattern <3pt, 1pt>
\putrule from 15.6 7.95 to 16.4 7.95   
\putrule from 15.6 9.15 to 16.4 9.15   
\putrule from 16.6 7.76 to 17.4 7.76   
\setsolid
\putrule from 16.6 8.19 to 17.4 8.19
\setdashpattern <3pt, 1pt>
\putrule from 17.6 9.16 to 18.4 9.16   
\putrule from 19.6 8.92 to 20.4 8.92   
\putrule from 20.6 10.38 to 21.4 10.38   
\putrule from 22.6 10.19 to 23.4 10.19   
\putrule from 23.6 11.66 to 24.4 11.66   
\setsolid
\linethickness=0.7pt
\putrule from 18 4 to 18 5.75
\putrule from 18 5.75 to 25 5.75
\putrule from 25 5.75 to 25 4
\putrule from 25 4 to 18 4
\linethickness=2pt
\putrule from 18.5 5.25 to 19.3 5.25
\put {Fit Masses} [l] at 19.5 5.25
\setdashpattern <3pt, 1pt>
\putrule from 18.5 4.6 to 19.3 4.6
\setsolid
\put {Predicted Masses} [l] at 19.5 4.6
\endpicture$$

\vfill\eject

$$\beginpicture
\setcoordinatesystem units <0.750truein,0.250truein>
\setplotarea x from -1 to 7, y from -31 to 2.0
%
%
\put {Table 6.  Masses and Mixing Angles of the $L=1$ Baryons
in the Quark Model with} [l] at -1 -32
\put {$SU(3)$ Symmetric Interactions.} [l] at -1 -32.75 
%
%
\put {Baryon} [c] at -0.25 0.75 
\put {Mass} [c] at 1 1.25 
\put {(Exp)} [c] at 1 0.375 
\put {Mass} [c] at 2 1.25
\put {(Fit)} [c] at 2 0.375
\put {$^4$8} [c] at 3 0.75
\put {$^2$8} [c] at 4 0.75
\put {$^2$10} [c] at 5 0.75
\put {$^2$1} [c] at 6 0.75
\linethickness=1pt
\putrule from -1 1.75 to 6.5 1.75
\putrule from -1 -31 to 6.5 -31
\linethickness=1.5pt
\putrule from -1 -0.125 to 6.5 -0.125
%
\put {$N$} [c] at -0.625 -1
\put {$J=\half$} [c] at 0 -1
\put {$1535^{+20}_{-15}$} [c] at 1 -1
\put {1522} [c] at 2 -1
\put {-0.2404} [c] at 3 -1
\put {-0.9707} [c] at 4 -1
\put {*} [c] at 5 -1
\put {*} [c] at 6 -1
\put {$1650^{+30}_{-10}$} [c] at 1 -2
\put {1647} [c] at 2 -2
\put {-0.9707} [c] at 3 -2
\put {0.2404} [c] at 4 -2
\put {*} [c] at 5 -2
\put {*} [c] at 6 -2
%
\put {$N$} [c] at -0.625 -3
\put {$J=\thalf$} [c] at 0 -3
\put {$1520^{+10}_{-5}$} [c] at 1 -3
\put {1503} [c] at 2 -3
\put {0.8903} [c] at 3 -3
\put {0.4554} [c] at 4 -3
\put {*} [c] at 5 -3
\put {*} [c] at 6 -3
\put {$1700^{+50}_{-50}$} [c] at 1 -4
\put {1584} [c] at 2 -4
\put {-0.4554} [c] at 3 -4
\put {0.8903} [c] at 4 -4
\put {*} [c] at 5 -4
\put {*} [c] at 6 -4
%
\put {$N$} [c] at -0.625 -5
\put {$J=\fhalf$} [c] at 0 -5
\put {$1675^{+10}_{-5}$} [c] at 1 -5
\put {1671} [c] at 2 -5
\put {1.0000} [c] at 3 -5
\put {*} [c] at 4 -5
\put {*} [c] at 5 -5
\put {*} [c] at 6 -5
%
\put {$\Delta$} [c] at -0.625 -6
\put {$J=\half$} [c] at 0 -6
\put {$1620^{+55}_{-5}$} [c] at 1 -6
\put {1761} [c] at 2 -6
\put {*} [c] at 3 -6
\put {*} [c] at 4 -6
\put {1.0000} [c] at 5 -6
\put {*} [c] at 6 -6
%
\put {$\Delta$} [c] at -0.625 -7
\put {$J=\thalf$} [c] at 0 -7
\put {$1700^{+70}_{-30}$} [c] at 1 -7
\put {1732} [c] at 2 -7
\put {*} [c] at 3 -7
\put {*} [c] at 4 -7
\put {1.0000} [c] at 5 -7
\put {*} [c] at 6 -7
%
\put {$\Sigma$} [c] at -0.625 -8
\put {$J=\half$} [c] at 0 -8
\put {**} [c] at 1 -8
\put {1639} [c] at 2 -8
\put {-0.2404} [c] at 3 -8
\put {-0.9707} [c] at 4 -8
\put {0.0000} [c] at 5 -8
\put {*} [c] at 6 -8
\put {$1750^{+50}_{-20}$} [c] at 1 -9
\put {1764} [c] at 2 -9
\put {-0.9707} [c] at 3 -9
\put {0.2404} [c] at 4 -9
\put {0.0000} [c] at 5 -9
\put {*} [c] at 6 -9
\put {**} [c] at 1 -10
\put {1878} [c] at 2 -10
\put {0.0000} [c] at 3 -10
\put {0.0000} [c] at 4 -10
\put {1.0000} [c] at 5 -10
\put {*} [c] at 6 -10
%
\put {$\Sigma$} [c] at -0.625 -11
\put {$J=\thalf$} [c] at 0 -11
\put {**} [c] at 1 -11
\put {1620} [c] at 2 -11
\put {0.8903} [c] at 3 -11
\put {0.4554} [c] at 4 -11
\put {0.0000} [c] at 5 -11
\put {*} [c] at 6 -11
\put {$1670^{+15}_{-5}$} [c] at 1 -12
\put {1701} [c] at 2 -12
\put {-0.4554} [c] at 3 -12
\put {0.8903} [c] at 4 -12
\put {0.9959} [c] at 5 -12
\put {*} [c] at 6 -12
\put {$1940^{+10}_{-40}$} [c] at 1 -13
\put {1849} [c] at 2 -13
\put {0.0000} [c] at 3 -13
\put {0.0000} [c] at 4 -13
\put {1.0000} [c] at 5 -13
\put {*} [c] at 6 -13
%
\put {$\Sigma$} [c] at -0.625 -14
\put {$J=\fhalf$} [c] at 0 -14
\put {$1775^{+5}_{-5}$} [c] at 1 -14
\put {1788} [c] at 2 -14
\put {1.0000} [c] at 3 -14
\put {*} [c] at 4 -14
\put {*} [c] at 5 -14
\put {*} [c] at 6 -14
%
\put {$\Lambda$} [c] at -0.625 -15
\put {$J=\half$} [c] at 0 -15
\put {$1407^{+4}_{-4}$} [c] at 1 -15
\put {1414} [c] at 2 -15
\put {0.0000} [c] at 3 -15
\put {0.0000} [c] at 4 -15
\put {*} [c] at 5 -15
\put {1.0000} [c] at 6 -15
\put {$1670^{+10}_{-10}$} [c] at 1 -16
\put {1639} [c] at 2 -16
\put {-0.2404} [c] at 3 -16
\put {-0.9707} [c] at 4 -16
\put {*} [c] at 5 -16
\put {0.0000} [c] at 6 -16
\put {$1800^{+50}_{-80}$} [c] at 1 -17
\put {1763} [c] at 2 -17
\put {-0.9707} [c] at 3 -17
\put {0.2404} [c] at 4 -17
\put {*} [c] at 5 -17
\put {0.0000} [c] at 6 -17
%
\put {$\Lambda$} [c] at -0.625 -18
\put {$J=\thalf$} [c] at 0 -18
\put {$1519.5^{+1}_{-1}$} [c] at 1 -18
\put {1519} [c] at 2 -18
\put {0.0000} [c] at 3 -18
\put {0.0000} [c] at 4 -18
\put {*} [c] at 5 -18
\put {1.0000} [c] at 6 -18
\put {**} [c] at 1 -19
\put {1620} [c] at 2 -19
\put {0.8903} [c] at 3 -19
\put {0.4554} [c] at 4 -19
\put {*} [c] at 5 -19
\put {0.0000} [c] at 6 -19
\put {$1690^{+5}_{-5}$} [c] at 1 -20
\put {1701} [c] at 2 -20
\put {-0.4554} [c] at 3 -20
\put {0.8903} [c] at 4 -20
\put {*} [c] at 5 -20
\put {0.0000} [c] at 6 -20
%
\put {$\Lambda$} [c] at -0.625 -21
\put {$J=\fhalf$} [c] at 0 -21
\put {$1830^{+0}_{-20}$} [c] at 1 -21
\put {1788} [c] at 2 -21
\put {1.0000} [c] at 3 -21
\put {*} [c] at 4 -21
\put {*} [c] at 5 -21
\put {*} [c] at 6 -21
%
\put {$\Xi$} [c] at -0.625 -22
\put {$J=\half$} [c] at 0 -22
\put {**} [c] at 1 -22
\put {1756} [c] at 2 -22
\put {-0.2404} [c] at 3 -22
\put {-0.9707} [c] at 4 -22
\put {*} [c] at 5 -22
\put {*} [c] at 6 -22
\put {**} [c] at 1 -23
\put {1880} [c] at 2 -23
\put {-0.9707} [c] at 3 -23
\put {0.2404} [c] at 4 -23
\put {*} [c] at 5 -23
\put {*} [c] at 6 -23
%
\put {$\Xi$} [c] at -0.625 -24
\put {$J=\thalf$} [c] at 0 -24
\put {**} [c] at 1 -24
\put {1736} [c] at 2 -24
\put {0.8903} [c] at 3 -24
\put {0.4554} [c] at 4 -24
\put {*} [c] at 5 -24
\put {*} [c] at 6 -24
\put {$1823^{+5}_{-5}$} [c] at 1 -25
\put {1818} [c] at 2 -25
\put {-0.4554} [c] at 3 -25
\put {0.8903} [c] at 4 -25
\put {*} [c] at 5 -25
\put {*} [c] at 6 -25
%
\put {$\Xi$} [c] at -0.625 -26
\put {$J=\fhalf$} [c] at 0 -26
\put {**} [c] at 1 -26
\put {1904} [c] at 2 -26
\put {1.0000} [c] at 3 -26
\put {*} [c] at 4 -26
\put {*} [c] at 5 -26
\put {*} [c] at 6 -26
%
\put {$\Xi^*$} [c] at -0.625 -27
\put {$J=\half$} [c] at 0 -27
\put {**} [c] at 1 -27
\put {1995} [c] at 2 -27
\put {*} [c] at 3 -27
\put {*} [c] at 4 -27
\put {1.0000} [c] at 5 -27
\put {*} [c] at 6 -27
%
\put {$\Xi^*$} [c] at -0.625 -28
\put {$J=\thalf$} [c] at 0 -28
\put {**} [c] at 1 -28
\put {1966} [c] at 2 -28
\put {*} [c] at 3 -28
\put {*} [c] at 4 -28
\put {1.0000} [c] at 5 -28
\put {*} [c] at 6 -28
%
\put {$\Omega$} [c] at -0.625 -29
\put {$J=\half$} [c] at 0 -29
\put {**} [c] at 1 -29
\put {2112} [c] at 2 -29
\put {*} [c] at 3 -29
\put {*} [c] at 4 -29
\put {1.0000} [c] at 5 -29
\put {*} [c] at 6 -29
%
\put {$\Omega$} [c] at -0.625 -30
\put {$J=\thalf$} [c] at 0 -30
\put {**} [c] at 1 -30
\put {2083} [c] at 2 -30
\put {*} [c] at 3 -30
\put {*} [c] at 4 -30
\put {1.0000} [c] at 5 -30
\put {*} [c] at 6 -30
\endpicture$$

\vfill\eject

$$\beginpicture
\setcoordinatesystem units <0.750truein,0.250truein>
\setplotarea x from -1 to 7, y from -31 to 2.0
%
%
\put {Table 7.  Masses and Mixing Angles of the $L=1$ Baryons 
in the Chiral Quark Model} [l] at -1 -32
\put {with $SU(3)$ Symmetric Interactions.} [l] at -1 -32.75 
%
%
\put {Baryon} [c] at -0.25 0.75 
\put {Mass} [c] at 1 1.25 
\put {(Exp)} [c] at 1 0.375 
\put {Mass} [c] at 2 1.25
\put {(Fit)} [c] at 2 0.375
\put {$^4$8} [c] at 3 0.75
\put {$^2$8} [c] at 4 0.75
\put {$^2$10} [c] at 5 0.75
\put {$^2$1} [c] at 6 0.75
\linethickness=1pt
\putrule from -1 1.75 to 6.5 1.75
\putrule from -1 -31 to 6.5 -31
\linethickness=1.5pt
\putrule from -1 -0.125 to 6.5 -0.125
%
\put {$N$} [c] at -0.625 -1
\put {$J=\half$} [c] at 0 -1
\put {$1535^{+20}_{-15}$} [c] at 1 -1
\put {1540} [c] at 2 -1
\put {0.6313} [c] at 3 -1
\put {-0.7755} [c] at 4 -1
\put {*} [c] at 5 -1
\put {*} [c] at 6 -1
\put {$1650^{+30}_{-10}$} [c] at 1 -2
\put {1660} [c] at 2 -2
\put {-0.7755} [c] at 3 -2
\put {-0.6313} [c] at 4 -2
\put {*} [c] at 5 -2
\put {*} [c] at 6 -2
%
\put {$N$} [c] at -0.625 -3
\put {$J=\thalf$} [c] at 0 -3
\put {$1520^{+10}_{-5}$} [c] at 1 -3
\put {1521} [c] at 2 -3
\put {-0.1808} [c] at 3 -3
\put {-0.9835} [c] at 4 -3
\put {*} [c] at 5 -3
\put {*} [c] at 6 -3
\put {$1700^{+50}_{-50}$} [c] at 1 -4
\put {1564} [c] at 2 -4
\put {-0.9835} [c] at 3 -4
\put {0.1808} [c] at 4 -4
\put {*} [c] at 5 -4
\put {*} [c] at 6 -4
%
\put {$N$} [c] at -0.625 -5
\put {$J=\fhalf$} [c] at 0 -5
\put {$1675^{+10}_{-5}$} [c] at 1 -5
\put {1661} [c] at 2 -5
\put {1.0000} [c] at 3 -5
\put {*} [c] at 4 -5
\put {*} [c] at 5 -5
\put {*} [c] at 6 -5
%
\put {$\Delta$} [c] at -0.625 -6
\put {$J=\half$} [c] at 0 -6
\put {$1620^{+55}_{-5}$} [c] at 1 -6
\put {1636} [c] at 2 -6
\put {*} [c] at 3 -6
\put {*} [c] at 4 -6
\put {1.0000} [c] at 5 -6
\put {*} [c] at 6 -6
%
\put {$\Delta$} [c] at -0.625 -7
\put {$J=\thalf$} [c] at 0 -7
\put {$1700^{+70}_{-30}$} [c] at 1 -7
\put {1783} [c] at 2 -7
\put {*} [c] at 3 -7
\put {*} [c] at 4 -7
\put {1.0000} [c] at 5 -7
\put {*} [c] at 6 -7
%
\put {$\Sigma$} [c] at -0.625 -8
\put {$J=\half$} [c] at 0 -8
\put {$(1620)$} [c] at 1 -8
\put {1668} [c] at 2 -8
\put {0.6313} [c] at 3 -8
\put {-0.7755} [c] at 4 -8
\put {0.0000} [c] at 5 -8
\put {*} [c] at 6 -8
\put {$1750^{+50}_{-20}$} [c] at 1 -9
\put {1764} [c] at 2 -9
\put {0.0000} [c] at 3 -9
\put {0.0000} [c] at 4 -9
\put {1.0000} [c] at 5 -9
\put {*} [c] at 6 -9
\put {**} [c] at 1 -10
\put {1788} [c] at 2 -10
\put {-0.7755} [c] at 3 -10
\put {-0.6313} [c] at 4 -10
\put {0.0000} [c] at 5 -10
\put {*} [c] at 6 -10
%
\put {$\Sigma$} [c] at -0.625 -11
\put {$J=\thalf$} [c] at 0 -11
\put {$(1580)$} [c] at 1 -11
\put {1648} [c] at 2 -11
\put {-0.1808} [c] at 3 -11
\put {-0.9835} [c] at 4 -11
\put {0.0000} [c] at 5 -11
\put {*} [c] at 6 -11
\put {$1670^{+15}_{-5}$} [c] at 1 -12
\put {1692} [c] at 2 -12
\put {-0.9835} [c] at 3 -12
\put {0.1808} [c] at 4 -12
\put {0.0000} [c] at 5 -12
\put {*} [c] at 6 -12
\put {$1940^{+10}_{-40}$} [c] at 1 -13
\put {1911} [c] at 2 -13
\put {0.0000} [c] at 3 -13
\put {0.0000} [c] at 4 -13
\put {1.0000} [c] at 5 -13
\put {*} [c] at 6 -13
%
\put {$\Sigma$} [c] at -0.625 -14
\put {$J=\fhalf$} [c] at 0 -14
\put {$1775^{+5}_{-5}$} [c] at 1 -14
\put {1788} [c] at 2 -14
\put {1.0000} [c] at 3 -14
\put {*} [c] at 4 -14
\put {*} [c] at 5 -14
\put {*} [c] at 6 -14
%
\put {$\Lambda$} [c] at -0.625 -15
\put {$J=\half$} [c] at 0 -15
\put {$1407^{+4}_{-4}$} [c] at 1 -15
\put {1407} [c] at 2 -15
\put {0.0000} [c] at 3 -15
\put {0.0000} [c] at 4 -15
\put {*} [c] at 5 -15
\put {1.0000} [c] at 6 -15
\put {$1670^{+10}_{-10}$} [c] at 1 -16
\put {1668} [c] at 2 -16
\put {0.6313} [c] at 3 -16
\put {-0.7755} [c] at 4 -16
\put {*} [c] at 5 -16
\put {0.0000} [c] at 6 -16
\put {$1800^{+50}_{-80}$} [c] at 1 -17
\put {1788} [c] at 2 -17
\put {-0.7755} [c] at 3 -17
\put {-0.6313} [c] at 4 -17
\put {*} [c] at 5 -17
\put {0.0000} [c] at 6 -17
%
\put {$\Lambda$} [c] at -0.625 -18
\put {$J=\thalf$} [c] at 0 -18
\put {$1519.5^{+1}_{-1}$} [c] at 1 -18
\put {1519} [c] at 2 -18
\put {0.0000} [c] at 3 -18
\put {0.0000} [c] at 4 -18
\put {*} [c] at 5 -18
\put {1.0000} [c] at 6 -18
\put {**} [c] at 1 -19
\put {1649} [c] at 2 -19
\put {-0.1808} [c] at 3 -19
\put {-0.9835} [c] at 4 -19
\put {*} [c] at 5 -19
\put {0.0000} [c] at 6 -19
\put {$1690^{+5}_{-5}$} [c] at 1 -20
\put {1692} [c] at 2 -20
\put {-0.9835} [c] at 3 -20
\put {0.1808} [c] at 4 -20
\put {*} [c] at 5 -20
\put {0.0000} [c] at 6 -20
%
\put {$\Lambda$} [c] at -0.625 -21
\put {$J=\fhalf$} [c] at 0 -21
\put {$1830^{+0}_{-20}$} [c] at 1 -21
\put {1788} [c] at 2 -21
\put {1.0000} [c] at 3 -21
\put {*} [c] at 4 -21
\put {*} [c] at 5 -21
\put {*} [c] at 6 -21
%
\put {$\Xi$} [c] at -0.625 -22
\put {$J=\half$} [c] at 0 -22
\put {**} [c] at 1 -22
\put {1795} [c] at 2 -22
\put {0.6313} [c] at 3 -22
\put {-0.7755} [c] at 4 -22
\put {*} [c] at 5 -22
\put {*} [c] at 6 -22
\put {**} [c] at 1 -23
\put {1915} [c] at 2 -23
\put {-0.7755} [c] at 3 -23
\put {-0.6313} [c] at 4 -23
\put {*} [c] at 5 -23
\put {*} [c] at 6 -23
%
\put {$\Xi$} [c] at -0.625 -24
\put {$J=\thalf$} [c] at 0 -24
\put {**} [c] at 1 -24
\put {1776} [c] at 2 -24
\put {-0.1808} [c] at 3 -24
\put {-0.9835} [c] at 4 -24
\put {*} [c] at 5 -24
\put {*} [c] at 6 -24
\put {$1823^{+5}_{-5}$} [c] at 1 -25
\put {1819} [c] at 2 -25
\put {-0.9835} [c] at 3 -25
\put {0.1808} [c] at 4 -25
\put {*} [c] at 5 -25
\put {*} [c] at 6 -25
%
\put {$\Xi$} [c] at -0.625 -26
\put {$J=\fhalf$} [c] at 0 -26
\put {**} [c] at 1 -26
\put {1916} [c] at 2 -26
\put {1.0000} [c] at 3 -26
\put {*} [c] at 4 -26
\put {*} [c] at 5 -26
\put {*} [c] at 6 -26
%
\put {$\Xi^*$} [c] at -0.625 -27
\put {$J=\half$} [c] at 0 -27
\put {**} [c] at 1 -27
\put {1892} [c] at 2 -27
\put {*} [c] at 3 -27
\put {*} [c] at 4 -27
\put {1.0000} [c] at 5 -27
\put {*} [c] at 6 -27
%
\put {$\Xi^*$} [c] at -0.625 -28
\put {$J=\thalf$} [c] at 0 -28
\put {**} [c] at 1 -28
\put {2038} [c] at 2 -28
\put {*} [c] at 3 -28
\put {*} [c] at 4 -28
\put {1.0000} [c] at 5 -28
\put {*} [c] at 6 -28
%
\put {$\Omega$} [c] at -0.625 -29
\put {$J=\half$} [c] at 0 -29
\put {**} [c] at 1 -29
\put {2019} [c] at 2 -29
\put {*} [c] at 3 -29
\put {*} [c] at 4 -29
\put {1.0000} [c] at 5 -29
\put {*} [c] at 6 -29
%
\put {$\Omega$} [c] at -0.625 -30
\put {$J=\thalf$} [c] at 0 -30
\put {**} [c] at 1 -30
\put {2166} [c] at 2 -30
\put {*} [c] at 3 -30
\put {*} [c] at 4 -30
\put {1.0000} [c] at 5 -30
\put {*} [c] at 6 -30
\endpicture$$

\bye